\documentclass[journal]{IEEEtran}
\IEEEoverridecommandlockouts
\usepackage{cite}
\usepackage{amsmath,amssymb,amsfonts}
\usepackage{bbm}
\usepackage{algorithm}
\usepackage[noend]{algpseudocode}
\usepackage{graphicx}
\usepackage{textcomp}
\usepackage{xcolor}
\usepackage{ulem}
\def\BibTeX{{\rm B\kern-.05em{\sc i\kern-.025em b}\kern-.08em
    T\kern-.1667em\lower.7ex\hbox{E}\kern-.125emX}}
\usepackage{float}
\usepackage{amsmath}
\newtheorem{theorem}{Theorem}
\newtheorem{lemma}{Lemma}
\newtheorem{definition}{Definition}

\newtheorem{corollary}{Corollary}

\allowdisplaybreaks
\newtheorem{pb}{Problem}

\newcommand{\revise}{}

\begin{document}

\title{Minimizing Age of Information with Power Constraints: Multi-User Opportunistic Scheduling in Multi-State Time-Varying Channels}
\author{Haoyue~Tang,~\IEEEmembership{Student Member,~IEEE},~Jintao Wang,~\IEEEmembership{Senior Member,~IEEE},~Linqi~Song,~\IEEEmembership{Member, IEEE},\\and~Jian~Song,~\IEEEmembership{Fellow,~IEEE}
	\thanks{Manuscript received July 1, 2019. This work was supported in part by the National Key R\&D Program of China under Grant 2017YFE0112300, Beijing National Research Center for Information Science and Technology under Grant BNR2019RC01014 and BNR2019TD01001, the Tsinghua University Tutor Research Fund and the Hong Kong RGC ECS grant 9048149. This work has been presented in part in 2019 57th Annual Allerton Conference on Communication, Control, and Computing (Allerton)\cite{haoyue}. (Corresponding author: Jintao Wang.)}	
	\thanks{H. Tang, J. Wang and J. Song are with the Department of Electronic Engineering, Tsinghua University, Beijing 100084, China and Beijing National Research Center for Information Science and Technology (BNRist). J. Wang and J. Song are also with Research Institute of Tsinghua University in Shenzhen, Shenzhen, 518057. (e-mail: tanghaoyue13@tsinghua.org.cn; wangjintao@tsinghua.edu.cn; jsong@tsinghua.edu.cn). L. Song is with the City University of Hong Kong, 83 Tat Chee Ave, Kowloon Tong, Hong Kong. (e-mail: linqi.song@cityu.edu.hk).}
}

\maketitle

\begin{abstract}
This work is motivated by the need of collecting fresh data from power-constrained sensors in the industrial Internet of Things (IIoT) network. A recently proposed metric, the \textit{Age of Information} (AoI) is adopted to measure data freshness from the perspective of the central controller in the IIoT network. We wonder what is the minimum average AoI the network can achieve and how to design scheduling algorithms to approach it. To answer these questions when the channel states of the network are Markov time-varying and scheduling decisions are restricted to bandwidth constraint, we first decouple the multi-sensor scheduling problem into a single-sensor constrained Markov decision process (CMDP) by relaxing the hard bandwidth constraint. 
Next we exploit the threshold structure of the optimal policy for the decoupled single sensor CMDP and obtain the optimum solution through linear programming (LP). 
Finally, an asymptotically optimal truncated policy that can satisfy the hard bandwidth constraint is built upon the optimal solution to each of the decoupled single-sensor. Our investigation shows that to obtain a small AoI performance: (1) The scheduler exploits good channels to schedule sensors supported by limited power; (2) Sensors equipped with enough transmission power are updated in a timely manner such that the bandwidth constraint can be satisfied.
\end{abstract}


\section{Introduction}
	The forthcoming Industrial 4.0 revolution brings more stringent data freshness requirement to support the higher level automated applications such as industrial manufacturing and factory automation \cite{iiot2}. In many of these applications, the monitor or the central controller collects data from sensors tracking real-time processes via time-varying wireless links \cite{iiot3}. The finite battery capacity, limited recharge resources \cite{iiot4} and wireless interference constraints cast restrictions on real time data sampling process and communications between the sensor and the monitor. In addition, data freshness requirement is different from traditional quality of service (QoS) guarantees such as communication latency and throughput. Thus, it is of great importance to revisit sampling and scheduling strategies in wireless networks in order to obtain more fresh information.
	
	Previous techniques on minimizing communication latency and maximizing throughput may not be applied directly to data freshness optimization, since low latency and high throughput may not fulfill a good data freshness requirement. A relevant metric that captures data freshness, the \textit{Age of Information} (AoI) \cite{yates12infocom}, namely the time elapsed since the generation time-stamp of the freshest information stored at the receiver, has received increasing attention. As have been shown in \cite{yates19tit,kam18tit,devassy19jsac}, analyzing AoI performance and guaranteeing low AoI requirement are especially challenging since the performance is affected by fundamental trade-off between communication throughput and transmission delay. 
	
	Moreover, combating the time-varying characteristic of wireless fading channels with limited communication resources such as power consumption and bandwidth is important but challenging in stochastic networks, since these constraints and randomness appear at different layers of the communication networks \cite{tony98tit} and require a joint design of physical and data link layer. In addition, the exponential growth of the cardinality of system states and action spaces, known as "\textit{the curse of dimension}", creates obstacles in searching for the optimal policy. 
	
\begin{figure}[h]
		\centering
		\includegraphics[width=.45\textwidth]{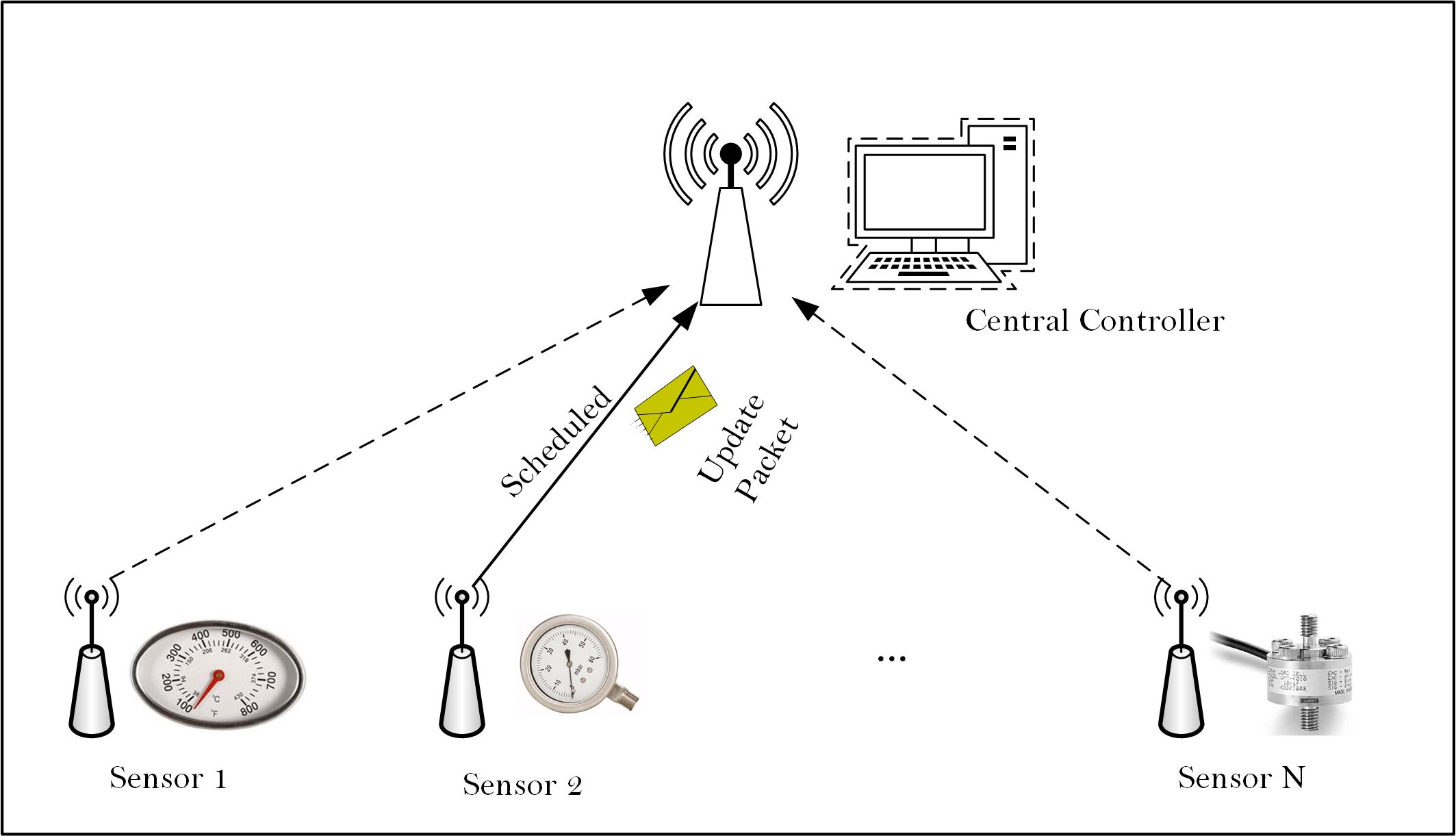}
		\caption{Illustration of a typical industrial Internet of Things (IIoT) network.}
		\label{networkmodel}
	\end{figure}

	To address these challenges, in our paper, we consider a single controller multi-sensor IIoT network where each sensor is scheduled to transmit update packet by the central controller, as depicted in Fig.~\ref{networkmodel}. The goal is to understand the how to design AoI minimization strategies in time-varying wireless channel with power constrained sensors. This scenario can be used to model the following applications in Industrial 4.0:  
		\begin{itemize}
		\item \textit{Factory Automation: }This application requires the central controller supervising all rounds of the production process in order to guarantee efficient and safe operation. Each sensor is charged by different amount of power and tracks different servers during the manufacturing process. The central controller designs efficient load balancing algorithm for parallel servers based on the current manufacturing process reported by each sensor.
		\item \textit{Intelligent Logistic: }\revise{The design of efficient intelligent logistic system requires precise observation and estimation of user demands.} In this scenario, sensors can be viewed as power constrained wireless hot spots that collect time-varying user preferences and requirements, while the central controller makes real-time scheduling decision in the logistic network based on these demands. 
	\end{itemize} 
	
	\revise{A main feature of the model is that the channels are multi-state time-varying} and information collected by the sensors are time sensitive. We generalize our previous work \cite{haoyue} by assuming the channel evolution has Markov properties, which is more suitable to capture real-time fading effect. To ensure successful transmission, different level of transmission power is used in different channel state, while each sensor has an average power consumption constraint. The overall objective is to design scheduling policy that meets both power and bandwidth constraint, while the expected average AoI over the entire network can be minimized. Based on a single sensor level decomposition through a relaxation of the hard bandwidth constraint, we propose a truncated scheduling policy that can achieve an asymptotic optimal average AoI performance over the entire network. 

The main contributions of the paper are summarized as follows:
\begin{itemize}
	\item  Consider that fresh update packet can be transmitted at every transmission, we propose a cross-layer framework to study AoI minimization scheduling in multi-user bandwidth limited network with power constrained sensors. The channel is modeled to be a finite-state ergodic Markov chain but remains constant in each slot. Different amount of transmit power is adopted in different channel state to ensure successful packet transmission. Unlike previous work, we consider both power and bandwidth constraint in a multi-user setup. This model captures key features of practical cross-layer network optimization problem and facilitates analysis.
	
	\item We decouple the multi-sensor scheduling problem into a single-sensor constrained Markov decision process (CMDP) by relaxing the hard bandwidth constraint and then through the Lagrange multiplier. The threshold structure of the optimal policy for the decoupled single-sensor CMDP is revealed, and the search for the optimal policy is converted into a Linear Programming (LP). \revise{This approach has not been used in AoI problems before.}
	
	\item We adopt a dual-method to search for the Lagrange multipliers such that the relaxed bandwidth constraint can be satisfied. Then, we propose an asymptotic optimum truncated scheduling policy so that the hard bandwidth constraint of the network can be satisfied. The performance of the algorithm is analyzed theoretically and verified through simulations. 
\end{itemize}

The remainder of this paper is organized as follows. We review some related work in Section II. The network model and the data freshness metric, AoI, are introduced in Section III. In Section IV, we decouple the multi-sensor scheduling problem into single-sensor level CMDP and search for the optimal policy through LP. In Section V, a truncated multi-sensor scheduling policy is proposed. Section VI evaluates and analyzes the performance of the proposed algorithm. Section VII draws the conclusion. 

\textit{Notations: }\revise{Vectors and matrices are written in boldface lower and upper letters, respectively.} The probability of event $\mathcal{A}$ given condition $\mathcal{B}$ is denoted as Pr$(\mathcal{A}|\mathcal{B})$. The expectation operation with regard to random variable $X$ is denoted as $\mathbb{E}_X[\cdot]$. The cardinality of a set $\Omega$ is denoted as $|\Omega|$.

\section{Related work}
	The analysis and optimization of AoI performance in average power constrained point to point communication system have been studied \cite{yates_15_isit,sun16infocom,sun17TIT,arafa_18_isit,yang_17_isit,ceran_18_wcnc,baknina_18_isit}. It is revealed that the optimal sampling policy with power constrained transmitter in the presence of queueing delay \cite{sun17TIT} and transmission failure \cite{ceran_18_wcnc} possesses a threshold structure, i.e., sampling and update transmission occur when information at the receiver is no longer fresh while the update packets, if successfully received, can significantly reduce data staleness. 
	
	Another line of work focuses on designing scheduling strategies to minimize AoI performance in multi-user wireless networks\cite{igor16allerton,igor_18_ton,talak2017allerton,talak_18_isit,talak_18_wiopt,tangisit,hsu17isit,jiang_isit_2018,lu_age_2018,igor18infocom}. When all the users in the network are identical and update packets can be generated at will, a greedy policy that schedules the user with the largest AoI is shown to be optimal \cite{igor16allerton}. When there is no packet-loss in the network, this greedy policy is equivalent to the round robin strategy, which is shown to be order optimal when update packets can not be generated at will and arrive randomly \cite{jiang_isit_2018}. In \cite{igor_18_ton}, it is revealed that users with relatively bad channel states are updated less frequently. \revise{Scheduling in networks with time-varying channels are studied in} \cite{talak_18_isit,talak_18_wiopt}, where channels with two states is considered, and centralized and decentralized policies to minimize AoI are proposed respectively.

	Cross-layer control strategy to minimize communication latency under transmit power constraints have been studied in \cite{borkar_18_tcns,chen_2018_ton,wang_17_GC,wang_19_tcomm,yang_10_twc,uysal_02_ton,berry_02_tit,singh_2019_tac}. In \cite{uysal_02_ton}, a Lazy scheduling policy that assigns scheduling decision based on the queue backlog is proposed. Considering the time-varying fading nature of wireless channels, rate and power adaptation strategy is proposed in \cite{berry_02_tit}. To minimize queueing delay in a point to point time-varying channel with average power constraint on the transmitter, a probabilistic scheduling strategy is proposed in \cite{wang_17_GC,wang_19_tcomm}. However the above work consider wireless fading to be an i.i.d process. When channel state evolution has Markov properties, scheduling to minimize delay performance and maximize throughput have been studied in \cite{chen_2018_ton,borkar_18_tcns,singh_2019_tac}. Scheduling policy based on value iteration is proposed in \cite{chen_2018_ton} and a Whittle-like index policy to achieve delay-power trade-off is studied in \cite{borkar_18_tcns}. In \cite{singh_2019_tac}, the multi-user power and bandwidth constrained scheduling problem is solved by packet level decomposition, and an asymptotically optimum truncated scheduling policy is proposed. Rajat \textit{et. al} studied a joint rate control and scheduling problem for age minimization under general interference constraints \cite{talak2017allerton}, where joint rate control and scheduling policies are investigated for age optimality, and a separation principle policy is found to be approximately optimal. However, no power constraint is considered in that work.
	
\section{System Model and Problem Formulation}
\subsection{Network Model}
We consider an industrial Internet of Things (IIoT) network as depicted in Fig.~\ref{networkmodel}, where a central controller collects time-sensitive data from $N$ sensors via wireless links. Let the time be slotted and use $t\in\{1, \cdots, T\}$ to denote the index of slot. Let the indicator function $u_n(t)=\{0, 1\}$ be a scheduling decision made by the central controller at the beginning of slot $t$. If $u_n(t)=1$, then sensor $n$ is scheduled to transmit update packet about his observation in slot $t$. We assume each successful transmission takes one slot and the packet will be received by the end of the slot. Due to limited bandwidth constraint, no more than $M$ sensors can be scheduled in each slot. We consider a non-trivial case and assume the bandwidth $M<N$, thus we have the following constraint on $u_n(t)$:
\begin{equation}
	\sum_{n=1}^Nu_n(t)\leq M, \forall t.
\end{equation}

To model the time-varying characteristic of the channel between each sensor and the central controller, we class each channel into $Q$ states and assume the channel state of sensor $n$, denoted by $\{q_n(t)\}$ is a $Q$-state ergodic Markov chain with transision probability $p^{(n)}_{i,j}\triangleq\text{Pr}(q_n(t+1)=j|q_n(t)=i)$.
If sensor $n$ is scheduled to transmit updates when the current channel state is $q$, \revise{in order to guarantee the channel capacity is larger than the size of an update packet}, it will consume $\omega(q)$ units of power. \revise{Similar to} \cite{uysal_02_ton,wang_19_tcomm,borkar_18_tcns}, \revise{we assume the transmitted packet will be successfully received by the central controller at the end of the slot.} For a typical scheduling decision $\mathbf{u}_n(\pi)=[u_n(1), \cdots, u_n(T)]$ of sensor $n$, the average power consumed in $T$ consecutive slots is:
\begin{equation}
E_n(\mathbf{u}_n(\pi))=\frac{1}{T}\sum_{t=1}^Tu_n(t)\omega(q_n(t)).
\end{equation}

\subsection{Age of Information}
We measure data freshness of the central controller by using the metric \textit{Age of Information} (AoI) \cite{yates12infocom}. By definition, the AoI is the time elapsed since the generation time-stamp of the freshest information at the receiver. An illustration of AoI evolution for a specific sensor is plotted in Fig. \ref{AoIillustration}:
\begin{figure}[h]
	\centering
	\includegraphics[width=.5\textwidth]{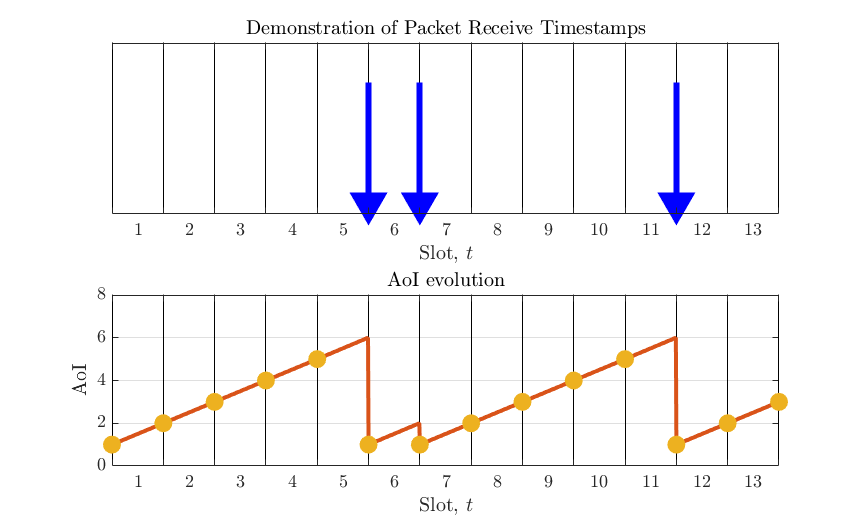}
	\caption{Illustration of AoI evolution of a specific sensor. On the top, sample sequence representing the receiving time-stamps of the generate-at-will update packets. On the bottom, sample paths of AoI (red). The yellow dots depict the AoI at the beginning of each slot. Upon receiving a new packet, the AoI will drop to $1$ at the beginning of next slot.}
	\label{AoIillustration}
\end{figure}

Let $x_n(t)$ be the AoI, i.e., the number of slots elapsed since the last delivery from sensor $n$ at the beginning of slot $t$. \revise{We consider a generate at will model similar to} \cite{yates_15_isit,igor_18_ton} \revise{and focus on minimizing the average AoI over the entire network. In this case, update packets generated before slot $t$ will be discarded and the system experiences no queueing delay. }Recall that if $u_n(t)=1$, sensor $n$ is scheduled in slot $t$ and an update containing the freshest information tracked by sensor $n$ will be received by the central controller, then by definition $x_n(t+1)=1$; otherwise, since there is no update packet received from sensor $n$ during slot $t$, $x_n(t)$ increases linearly and $x_n(t+1)=x_n(t)+1$. \revise{The AoI $x_n(t)$ evolves as follows:}
\begin{equation}
	x_n(t+1)=\begin{cases}
		1, &u_n(t)=1;\\
		x_n(t)+1, &u_n(t)=0. 
	\end{cases}
\end{equation}

\subsection{Problem Formulation}
For a given network setup with $N$ sensors and channel states evolution $\{p_{i,j}^{(n)}\}$, we measure the data freshness of the IIoT network by following policy $\pi$ in terms of the expected average AoI of all sensors at the beginning of each slot for a total of consecutive $T\rightarrow\infty$ slots, which can be computed as follows: 
\begin{align}
J(\pi)&=\lim_{T\rightarrow\infty}\{\frac{1}{NT}\mathbb{E}_{\pi}\left[\sum_{t=1}^T\sum_{n=1}^Nx_n(t)|\mathbf{x}(0)\right]\},
\end{align}where the vector $\mathbf{x}(t)=[x_1(t), x_2(t), \cdots, x_N(t)]\in\mathbb{N}^N$ denotes the AoI of all sensors at the beginning of slot $t$. In this work, we assume that all the sources have been synchronized initially, i.e., $\mathbf{x}(0)=\mathbf{1}$ and omit it henceforth.

Let $\Pi_{\text{NA}}$ denote the class of non-anticipated policies, i.e., scheduling decisions are made based on past, current AoIs $\{x_n(t)\}$, channel states $\{q_n(t)\}$ and their evolving probabilities $\{p_{i,j}^{(n)}\}$. No information about the future AoI or channel states can be used. We assume the average power constraint of each sensor is known by the central controller. In this research, we aim at designing policy $\pi\in\Pi_{\text{NA}}$ to minimize the average expected AoI of all the sensors, while the time average power consumption constraint of each sensor can be satisfied. The original bandwidth and power constrained AoI minimization problem (B\&P-Constrained AoI) is as follows:
\begin{pb}[B\&P-Constrained AoI]\label{pb:schedulingprimal}
\begin{subequations}
	\begin{align}
	\pi^*=\arg&\min_{\pi\in\Pi_{\text{NA}}}\lim\limits_{T\rightarrow\infty}\{\frac{1}{NT}\mathbb{E}_{\pi}\left[\sum_{t=1}^T\sum_{n=1}^Nx_n(t)\right]\},\label{objprimal}\\
	\text{s.t. }&\mathbb{E}_\pi\left[\sum_{n=1}^N u_n(t)\right]\leq M, \forall t,\label{hardinterference}\\
	&\lim_{T\rightarrow\infty}\frac{1}{T}\mathbb{E}_{\pi}\left[\sum_{t=1}^Tu_n(t)\omega(q_n(t))\right]\leq\mathcal{E}_n, \!\forall n\label{energyconstraint}. 
	\end{align}
\end{subequations}
\end{pb}

Notice that the hard bandwidth constraint \eqref{hardinterference} in every slot $t$ suggests, \revise{there are $\left(\begin{matrix}
	N\\1
	\end{matrix}\right)+\cdots+\left(\begin{matrix}
	N\\M
	\end{matrix}\right)$ possible scheduling decisions in each slot, it is hard to approach this problem through dynamic programming.} We tackle with this challenge through the following approaches:
\begin{itemize}
\item Inspired by \cite{singh_2019_tac,chen_2018_ton,yates17isit}, in Section IV-(A), we first relax the hard bandwidth constraint \eqref{hardinterference} and adopt a sensor level decomposition by using Lagrange multiplier. After relaxation, multiple sensors can be scheduled simultaneously. \item In Section V, we propose a truncated scheduling policy to satisfy the hard bandwidth constraint \eqref{hardinterference} based on the solution to each of the decoupled single sensor.
\end{itemize}
\section{Scheduling by Sensor-level decomposition}
In this section, we start by relaxing and decoupling the \textit{B\&P-Constrained AoI}, then formulate the decoupled single sensor scheduling problem into a constrained Markov decision process (CMDP). We exploit the threshold structure of the optimal stationary randomized policy and the optimal solution is solved through linear programming (LP). 
\subsection{Sensor Level Decomposition}
Let us first relax the hard constraint 
\eqref{hardinterference} into an time-average constraint, the relaxed bandwidth and power constrained AoI minimization problem (RB\&P-Constrained AoI) can be organized as follows:
\begin{pb}[RB\&P-Constrained AoI]
\begin{subequations}
	\begin{align}
	\pi_R^*=\arg&\min_{\pi\in\Pi_{\text{NA}}}\lim\limits_{T\rightarrow\infty}\{\frac{1}{NT}\mathbb{E}_{\pi}\left[\sum_{t=1}^T\sum_{n=1}^Nx_n(t)\right]\},\label{objrelaxed}\\
	\text{s.t. }&\lim_{T\rightarrow\infty}\mathbb{E}_\pi\left[\frac{1}{T}\sum_{t=1}^T\sum_{n=1}^N u_n(t)\right]\leq M,\label{relaxedinterference}\\
	&\lim_{T\rightarrow\infty}\frac{1}{T}\mathbb{E}_{\pi}\left[\sum_{t=1}^Tu_n(t)\omega(q_n(t))\right]\leq\mathcal{E}_n, \!\forall n. 
	\end{align}
	\label{relaxed}
	\end{subequations}
\end{pb}
\revise{Notice that any policy $\pi$ that satisfies the bandwidth constraint in the \textit{B\&P-Constrained AoI} satisfies the bandwidth constraint in \textit{RB\&P-Constrained AoI}, hence the average AoI obtained by $\pi_R^*$ formulates a lower bound on the average AoI obtained by $\pi^*$. To solve Problem 2, let us place the relaxed constraint into the objective function:} 
\begin{align}
	\mathcal{L}&(\pi, W)=\label{relaxedopt}\\
	&\lim_{T\rightarrow\infty}	\{\frac{1}{NT}\mathbb{E}_\pi\left[\sum_{n=1}^N\sum_{t=1}^T\left(x_n(t)	
	+Wu_n(t)-\frac{WM}{N}\right)\right]\}.\nonumber
\end{align}

For fixed multiplier $W$, denote $\pi(W)$ be the optimum policy that minimizes the Lagrange function Eq.~\eqref{relaxedopt}, i.e., \begin{equation}\pi(W)=\arg\min_{\pi\in\Pi_{\text{NA}}}\mathcal{L}(\pi, W).\label{eq:lagrange}
\end{equation}

Notice that the optimum policy $\pi_R^*$ to Problem 2 is a mixture of no more than two policies $\pi(W_1)$ and $\pi(W_2)$, which minimizes the Lagrange function under different multipliers $W_1$ and $W_2$, respectively. Thus, in the following analysis, we will first solve $\pi(W)$ for fixed $W$ and then provide how to obtain the two policies $\pi(W_1)$ and $\pi(W_2)$. 

To obtain policy $\pi(W)$ for fixed $W$, notice that the Lagrange multiplier $W\geq 0$ associates with the relaxed constraint can be viewed as a penalty incurred by policies that want to schedule more users than the relaxed constraint. For fixed $W$, the optimization problem \eqref{relaxedopt} can then be decoupled into $N$ single sensor AoI and scheduling penalty minimization problem with average power consumption constraint \eqref{energyconstraint}, then the decoupled single sensor power constrained cost minimization problem (\textit{Decoupled P-Constrained Cost}) can be written out as follows: 
\begin{pb}[Decoupled P-Constrained Cost]
\begin{subequations}
\begin{align}
	\pi_{d, n}^*&=\arg\min_{\pi\in\Pi_{\text{NA}}}\mathcal{L}(\pi_n, W), \text{where }\\
	\mathcal{L}_n(\pi_n, W)&=
	\lim_{T\rightarrow\infty}\frac{1}{T}\mathbb{E}_{\pi_n}\left[\sum_{t=1}^Tx_n(t)+Wu_n(t)\right],
	\label{singleopt}\\
	\text{s.t}&\lim_{T\rightarrow\infty}\frac{1}{T}\mathbb{E}_{\pi_n}\left[\sum_{t=1}^Tu_n(t)\omega(q_n(t))\right]\leq\mathcal{E}_n.
\end{align}
\end{subequations}
\end{pb} 

Since the primal relaxed problem \eqref{relaxedopt} gets decoupled, we omit the subscript $n$ henceforth. We formulate the \textit{Decoupled P-Constrained Cost} minimization problem into an CMDP in Section III-(B) and analyze the structure of the optimum policy in Section III-(C). In Section III-(D), we convert the single-sensor optimization problem with fixed $W$ into a Linear Programming (LP). 

\subsection{Constrained Markov Decision Process Formulation}
The decoupled single-sensor scheduling problem can be formulated into a CMDP that consists of a quadruplet $(\mathbb{S}, \mathbb{A}, \text{Pr}(\cdot|\cdot), C(\cdot, \cdot))$, each item is explained as follows:
\begin{itemize}
	\item \textbf{State Space: }The state of a sensor in slot $t$ is the current AoI and the channel state $(x(t), q(t))$. The state space $\mathbb{S}=\{x\times q\}$ is thus countable but infinite.
	\item \textbf{Action Space: }There are two possible actions $s\in\mathbb{A}=\{0, 1\}$, while $s(t)=1$ denotes the sensor is scheduled to deliver updates to the central controller in slot $t$, while $s(t)=0$ represents that the sensor keeps idle and is not scheduled. Notice that $s(t)$ is different from scheduling decision $u(t)$, which has strict bandwidth constraint.
	\item \textbf{Probability Transfer Function: }If the sensor is not scheduled during slot $t$, i.e., $s(t)=0$, then $x(t+1)=x(t)+1$, otherwise if the sensor is scheduled, then the AoI drops to $x(t+1)=1$. The channel state $q(t+1)$ evolves independently of $x(t)$ and only relies on $q(t)$ due to its Markov property, hence the probability transfer function from state $(x, q)$ is organized as follows:
\begin{equation}
	\text{Pr}((x,q)\rightarrow(x', q'))=\begin{cases}
	p_{q, q'},&\begin{matrix}\{s=0, x'=x+1\}\\
	\text{ or }\{s=1, x'=1\}\end{matrix};\\
	0,&\text{otherwise}.
	\end{cases}
\end{equation}
	\item \textbf{One-Step Cost: }For given state $(x, q)$, the one-step cost by taking action $s$ contains AoI growth and scheduling penalty, which can be computed as follows:
	\begin{subequations}
	\begin{equation}
		C_X(x, q, s)=x+Ws,
		\label{onestepcost}
	\end{equation}
	while the one-step power consumption is:
	\begin{equation}
		C_Q(x, q, s)=\omega(q)s.
	\end{equation}
\end{subequations}
\end{itemize}

The objective of the decoupled CMDP is to design a scheduling policy $\pi$ such that the following average cost over infinite horizon can be minimized: \[\lim_{T\rightarrow\infty}\frac{1}{T}\mathbb{E}_\pi\left[\sum_{t=1}^TC_X(x(t), q(t), s(t))\right],\] 
while the average power constraint is satisfied, \[\lim_{T\rightarrow\infty}\frac{1}{T}\mathbb{E}_\pi\left[\sum_{t=1}^TC_Q(x(t), q(t), s(t))\right]\leq \mathcal{E}.\] 
\subsection{Characterization of the Optimal Policy}
In this part, we focus on exploiting the threshold structure of the optimal policy. Before moving on, first we provide the formal definition of stationary randomized policies and stationary deterministic policies:

\begin{definition}
	Let $\Pi_{\text{SR}}$ and $\Pi_{\text{SD}}$ denote the class of stationary randomized and stationary deterministic policies, respectively. Given observation $(x(t)=x, q(t)=q)$, a stationary randomized policy $\pi_{\text{SR}}\in\Pi_{\text{SR}}$ chooses action $s(t)=1$ with probability measure $\xi_{x, q}\in[0, 1]$ for all $t$. A stationary deterministic policy $\pi_{\text{SD}}\in\Pi_{\text{SD}}$ selects action $s(t)=a(x, q)$, where $a(\cdot):(x, q)\rightarrow\{0,1\}$ is a deterministic mapping from state space to action space. 
\end{definition}

According to \cite[Theorem 4.4]{altman1999constrained}, the optimal policy to the above CMDP (\textit{Decoupled P-Constrained Cost}) has the following property:
\begin{corollary}
	An optimal stationary randomized policy $\pi_d^*\in\Pi_{\text{SR}}$ exists for the decoupled single sensor power constrained scheduling problem \eqref{singleopt}, and it is a mixture of no more than two stationary deterministic policies $\pi_{\text{SD1}}, \pi_{\text{SD2}}\in\Pi_{\text{SD}}$. Let $\rho$ be the weight of following stationary deterministic policy $\pi_{\text{SD1}}$ and $(1-\rho)$ be the weight of following $\pi_{\text{SD2}}$. Then the optimum policy is:
	\begin{equation}
	\pi_d^*=\rho\pi_{\text{SD1}}+(1-\rho)\pi_{\text{SD2}}.
	\label{mixedstrategy}
	\end{equation}
\end{corollary}
\begin{IEEEproof}
	\revise{According to} \cite[Theorem 6.3]{altman1999constrained} \revise{an optimum stationary randomized policy exists for constrained Markov decision process with infinite state and action space. Since the Lagrange relaxation remove only one constraint, according to} \cite[Theorem 4.4]{altman1999constrained}, the optimum policy is a mixture of two policies that minimize the Lagrange function with different multipliers $\lambda_1$ and $\lambda_2$. Such derivations is used similarly in \cite{ceran_18_wcnc}.
\end{IEEEproof}

To obtain the two deterministic policies $\pi_{\text{SD1}}$ and $\pi_{\text{SD2}}$, next we establish an unconstrained MDP by placing the average power consumption constraint into the objective function. Let $\lambda\geq 0$ be the Lagrange multiplier related to the average power constraint, we write out the Lagrange function and the goal of the unconstrained MDP is to minimize the following overall average cost (we omit the constant item $-\lambda\mathcal{E}$):
\begin{align}
\lim_{T\rightarrow\infty}\frac{1}{T}\mathbb{E}_\pi\left[\sum_{t=1}^T\left[C_X(x(t), q(t), s(t))\!+\!\lambda C_Q(x(t), q(t), s(t))\right]\right].
\label{unconstrainedMDP}
\end{align}

For given Lagrange multiplier $\lambda$, a stationary deterministic policy to minimize the above unconstrained cost exists. Denote $\gamma$ be the time-average cost by following the optimum strategy. Then, there exits a differential cost-to-go function $V(x, q)$ that satisfies the following Bellman equation:
\begin{align}
V(x, q)&+\gamma=\min\{C_X(x, q, 0)\!+\!\sum_{q'\!=\!1}^Qp_{q,q'}V(x\!+\!1, q'),\notag\\ &C_X(x, q, 1)\!+\!\sum_{q'\!=\!1}^Qp_{q,q'}V(1, q')\!+\!\lambda C_Q(x, q, 1)\},
\label{Bellman}
\end{align}
where $\gamma$ is the average cost by following the optimal policy. Next, we will prove the threshold structure of the stationary deterministic policy for given $\lambda$, which will present insight for the structure of the optimal stationary randomized policy to solve the \textit{Decoupled P-Constrained Cost} minimization problem.
\begin{lemma}
	With fixed $\lambda$, the optimal stationary deterministic policy for solving the \textit{Decoupled P-Constrained Cost} problem \eqref{unconstrainedMDP} possesses a threshold structure. That is there exists a sequence of threshold $\tau_q$ for each state, when $x\geq\tau_q$, the optimal action $s^*(x, q)=1$ and when $x<\tau_q$, $s^*(x, q)=0$. 
\end{lemma}

\begin{IEEEproof}
	The proof is provided in Appendix A. Here we provide an intuitive analysis. Since communication between the sensor and the controller is power constrained, we only schedule when the information is no longer fresh or the channel state is good, i.e., $x$ is large or $q$ is small. This behavior characterizes a threshold structure. 
\end{IEEEproof}

Notice that optimal stationary randomized policy $\pi_d^*$ to the CMDP \eqref{singleopt} is a randomization between no more than two stationary deterministic policies\cite{altman1999constrained}, each of them can be obtained by solving the unconstrained MDP \eqref{unconstrainedMDP} which possesses a threshold structure. Then it can be concluded there exists a set of thresholds $\tau_q$, for each state $(x, q)$, if $x\geq\max_q\tau_q$, the stationary randomized policy $\pi_d^*$ schedules the sensor.
\subsection{Probabilistic Scheduling Policy for Single Sensor Case}
Let us now investigate into the class of stationary randomized policies. Denote $\xi_{x, q}$ to be the probability that the sensor is scheduled to send updates in state $(x, q)$. We aim at finding a set of optimal transmission probability $\{\xi_{x, q}^*\}$ to solve the \textit{Decoupled P-Constrained Cost} problem. From Section IV-(C), since there exists a set of thresholds $\tau_q$, for each state $(x, q)$, if $x\geq\max_q\tau_q$, the stationary randomized policy is to schedule the sensor. Thus, the optimum policy $\pi_d^*$ must satisfy $\xi^*_{x, q}=1, \forall (x, q), x\geq\max_q\tau_q$. Therefore, for each of the decoupled single sensor problem, the AoI $x$ cannot be larger than the largest threshold $\max_{q}\tau_q$. To find the optimal policy, we choose a large bound $X_{\text{max}}$ for $x$ that can guarantee $X_\text{max}\geq \max_q\tau_q$. We only consider policy that satisfies $\xi_{x, q}=1, \forall x\geq X_\text{max}$ in the following analysis, since policies that do not have such properties are not optimum and thus can be excluded from the discussions.  

\begin{figure*}[htb]
	\centering
	\includegraphics[width=.7\textwidth]{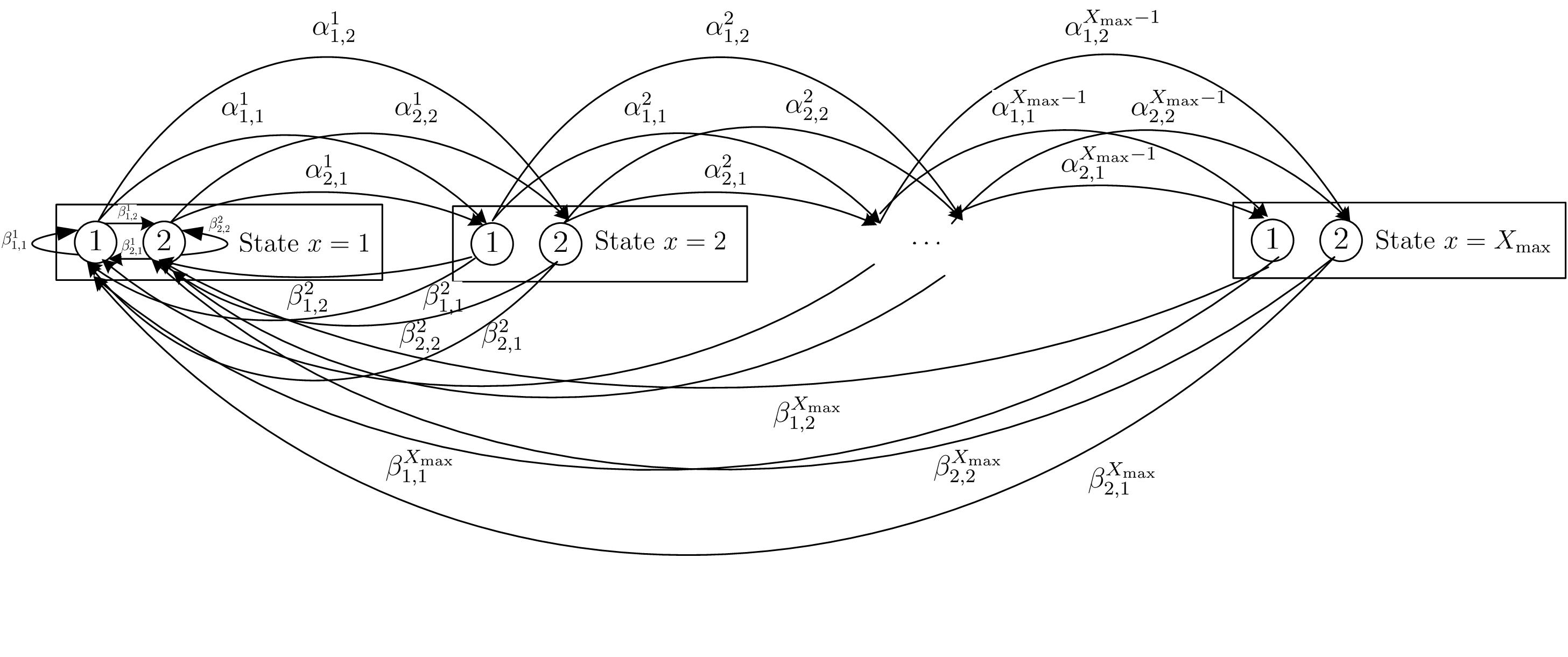}
	\caption{Illustrative of the probability transfer graph for a stationary randomized policy with $Q=2$ channel states. The circles denote channel state $q$ and the rectangles denote the sensor's AoI $x$. The forward state transmission probability $(x,q)$ to $(x+1,q')$ is $\alpha^x_{q, q'}$ and the backward state transmission probability from $(x,q)$ to $(1,q')$ is $\beta^x_{q,q'}$. }
	\label{probabilitytransfer}
\end{figure*}

Let $\mu_{x, q}$ denote the probability that the sensor's AoI is $x$ and the current channel state is $q$. To illustrate the state transition relationship, we provide transfer graph for $Q=2$ as an example in Fig. \ref{probabilitytransfer}. Let $\alpha_{q, q'}^x$ denote the one step forward state transition probability from $(x, q)$ to $(x+1, q')$ and let $\beta_{q, q'}^x$ be the backward transition probability from $(x, q)$ to $(1, q')$, respectively. From the discussed threshold structure of the stationary deterministic policies, with properly selected $X_{\text{max}}$, under the optimal scheduling policy, the steady state distribution $\mu_{X_{\text{max}}+1, q}=0,\forall q$. According to the probability transfer graph Fig. \ref{probabilitytransfer}, the forward and backward transition probability for a scheduling policy $\xi_{x, q}$ can be computed as follows:
\begin{subequations}
\begin{align}
\alpha_{q, q'}^x&=\text{Pr}((x, q)\rightarrow(x+1, q'))=(1-\xi_{x, q})p_{q, q'},\label{forward}\\ 
\beta^x_{q, q'}&=\text{Pr}((x,q)\rightarrow (1, q'))=\xi_{x, q}p_{q, q'}.\label{backward}
\end{align}
\end{subequations}

Let $\boldsymbol{\mu}=[\mu_{1, 1},\cdots,\mu_{1,Q},\cdots, \mu_{X_{\text{max}},1}, \cdots,\mu_{X_{\text{max}},Q}]^T$ be the steady state distribution. Let $\mathbf{Q}$ be the probability transfer matrix between the states, according to Fig. \ref{probabilitytransfer}, $\mathbf{Q}$ can be constructed as follows:
\begin{equation}
	\mathbf{Q}=\left[\begin{matrix}
	\boldsymbol{\beta}^1&\boldsymbol{\beta}^2&\cdots&\boldsymbol{\beta}^{X_\text{max}-1}&\boldsymbol{\beta}^{X_\text{max}}\\
	\boldsymbol{\alpha}^1&\boldsymbol{0}_Q&\cdots&\boldsymbol{0}_Q&\boldsymbol{0}_Q\\
	\boldsymbol{0}_Q&\boldsymbol{\alpha}^2&\cdots&\boldsymbol{0}_Q&\boldsymbol{0}_Q\\
	\cdots&\cdots&\cdots&\cdots\cdots\\
	\boldsymbol{0}_Q&\boldsymbol{0}_Q&\cdots&\boldsymbol{\alpha}^{X_\text{max}-1}&\boldsymbol{0}_Q\\
	\end{matrix}\right],
\end{equation} 
where vector $\mathbf{0}_{Q}$ is a $Q$-dimension vector with all the elements being 0. Matrices $\boldsymbol{\alpha}^x$ and $\boldsymbol{\beta}^x$ are the forward and backward transition matrix from state $x$, respectively, which can be computed as follows:
\begin{subequations}
	\begin{align}
		\boldsymbol{\alpha}^x&=\left[\begin{matrix}
		\alpha_{1, 1}^x&\alpha_{2, 1}^x&\cdots&\alpha_{Q,1}^x\\
		\alpha_{1, 2}^x&\alpha_{2, 2}^x&\cdots&\alpha_{Q,2}^x\\
		\cdots&\cdots&\cdots&\cdots\\
		\alpha_{1, Q}^x&\alpha_{2, Q}^x&\cdots&\alpha_{Q,Q}^x\\
		\end{matrix}\right],\\
		\boldsymbol{\beta}^x&=\left[\begin{matrix}
		\beta_{1, 1}^x&\beta_{2, 1}^x&\cdots&\beta_{Q,1}^x\\
		\beta_{1, 2}^x&\beta_{2, 2}^x&\cdots&\beta_{Q,2}^x\\
		\cdots&\cdots&\cdots&\cdots\\
		\beta_{1, Q}^x&\beta_{2, Q}^x&\cdots&\beta_{Q,Q}^x\\
		\end{matrix}\right].
	\end{align}
\end{subequations}

According to property of the steady state distribution, we have $\mathbf{Q}\boldsymbol{\mu}=\boldsymbol{\mu}$. In addition, considering that $\forall x\geq X_{\text{max}}+1$, the steady state distribution $\mu_{x, q}=0, \forall q$. We then have $\sum_{x=1}^{X_\text{max}}\sum_{q=1}^Q\mu_{x,q}=1$. Thus, the steady distribution $\boldsymbol{\mu}$ relates to strategy $\{\xi_{x, q}\}$ is the solution to the following linear equations:
\begin{equation}
\left[\begin{matrix}
\mathbf{Q}-\mathbf{I}_{Q X_{\text{max}}}\\
\boldsymbol{1}^T_{Q X_\text{max}}
\end{matrix}\right]\boldsymbol{\mu}=\left[\begin{matrix}
\boldsymbol{0}_{Q X_{\text{max}}}\\1
\end{matrix}\right],
\label{steadystatedistribution}
\end{equation}
where $\mathbf{1}_{Q X_{\text{max}}}$ is a $(Q\times X_{\text{max}})$-dimension column vector with all the elements being 1 and $\mathbf{I}_{Q X_\text{max}}$ is a $(Q\times X_{\text{max}})$ dimension identity matrix. 


Next, we will convert the search for the optimal stationary randomized scheduling strategy into an LP. We introduce a new set of variables $y_{x, q}=\mu_{x,q}\xi_{x, q}$, each denotes the probability of the sensor being in state $(x, q)$ and is scheduled to transmit an update. With this set of variables, we present the following theorem:
\begin{theorem}
	Solving the \textit{Decoupled P-Consrained Cost} minimization problem is equivalent to solve the following LP problem:
\begin{subequations}
	\begin{align}
	\{\mu_{x, q}^*,y_{x, q}^*\}&=\!\arg\min_{\{\mu_{x,q},y_{x, q}\}}\sum_{x=1}^{X_\text{max}}\sum_{q=1}^Q(Wy_{x, q}\!+\!x\mu_{x,q}),\label{stationaryLPobj}\\
	\text{s.t.  }	&\mu_{1,q}=\sum_{x=1}^X\sum_{q'=1}^Qy_{x, q'}p_{q',q},\label{stationaryLP1}\\
	&\mu_{x, q}= \sum_{q'=1}^Q(\mu_{x-1, q'}-y_{x-1,q'})p_{q',q}, \label{stationaryLP2}\\
	&\sum_{x=1}^{X_\text{max}}\sum_{q=1}^Q\mu_{x,q}=1, \label{stationaryLP3}\\
	&y_{x, q}\leq \mu_{x,q}, \label{stationaryLP4}\\
	&\sum_{x=1}^{X_\text{max}}\sum_{q=1}^{Q}y_{x, q}\omega(q)\leq\mathcal{E}\label{stationaryLP5}\\
	&0\leq\mu_{x,q}\leq 1, 0\leq y_{x, q}\leq 1, \forall x, q.\label{stationaryLP6} 
	\end{align}
\label{LP}
\end{subequations}
\end{theorem}
\begin{IEEEproof}
	Let us compute the equivalent time average cost to Eq.~\eqref{singleopt} as a sum of $\{\mu_{x, q}\}$ and $\{y_{x, q}\}$. The probability that the sensor is in state $(x, q)$ is $\mu_{x,q}$. With probability $\xi_{x, q}$, the sensor is selected to be scheduled and incurs a cost of $C_X(x, q, 1)=x+W$, and the sensor is selected to keep idle with probability $1-\xi_{x, q}$ and incurs a cost of $C_X(x, q, 0)=x$. Then the time average cost by following policy $\{\xi_{x, q}\}$ can be computed by:
\begin{align}
	&\sum_{x=1}^{X_\text{max}}\sum_{q=1}^Q\mu_{x,q}(\xi_{x, q}(x+W)+(1-\xi_{x, q})x)\nonumber\\
	=&\sum_{x=1}^{X_\text{max}}\sum_{q=1}^Q(Wy_{x, q}+x\mu_{x,q}). 
\end{align}

If the sensor is scheduled to transmit in state $(x, q)$, the power consumed is $\omega(q)$. Then, the time-average power consumed by employing policy $\{\xi_{x, q}\}$ is:
\begin{equation}
	\sum_{x=1}^{X_\text{max}}\sum_{q=1}^Q\mu_{x,q}\xi_{x, q}\omega(q)=\sum_{x=1}^{X_\text{max}}\sum_{q=1}^Qy_{x, q}\omega(q).
\end{equation}
With this equation the power constraint \eqref{energyconstraint} can be converted in the linear constraint \eqref{stationaryLP5}. The constraint Eq.~\eqref{stationaryLP1}-\eqref{stationaryLP3} can be obtained by substituting $\xi_{x, q}$ with $y_{x, q}$ and $\mu_{x, q}$ with relationship \eqref{steadystatedistribution}. Notice that $\xi_{x, q}\leq 1$, the inequality constraint \eqref{stationaryLP4} can be obtained. 
\end{IEEEproof}

Till now, we construct an LP problem to obtain $\mu_x$ and $y_{x, q}$ by following the optimum stationary randomized policy that minimizes the total cost with fixed Lagrange multiplier $W$. Next, the optimal stationary randomized scheduling policy to minimize Lagrange function Eq.~\eqref{singleopt} can be obtained through the relationship between $\xi_{x, q}, \mu_x$ and $y_{x, q}$. According to the threshold structure of each deterministic policy and Eq.~\eqref{mixedstrategy}, we will have the following property on $\xi^*_{x, q}$:
\begin{corollary}
For any channel state $q$, the optimal scheduling decisions $\xi_{\cdot, q}^*$ is monotonically increasing, i.e.,
			\begin{equation}
			\xi_{x_1, q}^*\leq \xi_{x_2, q}^*, \forall 1\leq x_1<x_2.
			\end{equation}
\end{corollary}

\section{Multi-sensor Opportunistic Scheduling}

In this section, we will provide an algorithm to determine the multiplier $W$ such that relaxed bandwidth constraint can be satisfied and \textit{RB\&P-Constrained AoI} problem can be solved. Then, we propose a truncated scheduling algorithm for the multi-sensor case that satisfies the original hard bandwidth constraint Eq.~\eqref{hardinterference}. 
\subsection{Determination of Lagrange Multiplier}  

Let $g(W)$ denote the Lagrange dual function, i.e., 
\begin{equation}
g(W)=\min_{\pi\in\Pi_{\text{NA}}}\mathcal{L}(\pi, W).\label{eq:langdual}
\end{equation}

Since the relaxed problem gets decoupled into $N$ single user CMDP, the dual function can be computed by:
\begin{align}
g(W)=\frac{1}{N}\sum_{n=1}^N&g_n(W)-WM, \text{where }\notag\\g_n(W)=\min_{\pi_n\in\Pi_{\text{NA}}}&\left(\mathcal{L}_n(\pi_n, W)\right), \text{ s.t. } \text{Eq.~(7c)}.
\end{align}

By Theorem 1, the CMDP that minimizes $\mathcal{L}_n(\pi, W)$ is equivalent to an LP, then $g_n(W)$ equals the average cost of the CMDP. Let $\overline{X}_n(W)$ and $\overline{A}_n(W)$ denote the average AoI and the average scheduling probability of sensor $n$, respectively. Let $\{y_{x, q}^{n, W}\}$ be the solution of sensor $n$' LP problem \eqref{LP} with multiplier $W$, function $g_n(W)$ can be computed as follows:
\begin{subequations}
	\begin{align}
	g_n(W)=\overline{X}_n(W)&+W\overline{A}_n(W),\\
	\text{where } \overline{X}_n(W)&=\sum_{x=1}^{X_{\text{max}}}\sum_{q=1}^Qx\mu_{x,q}^{n, W},\\
	\overline{A}_n(W)&=\sum_{x=1}^{X_{\text{max}}}\sum_{q=1}^Qy_{x, q}^{n, W}.
	\end{align}
\end{subequations}

According to \cite{mixing}, let $W^*$ be the supreme Lagrange multiplier such that policy $\pi(W)$ that minimizes the Lagrange function Eq.~\eqref{eq:lagrange} satisfies the relaxed bandwidth constraint, i.e., 
\[W^*=\sup\{W|\sum_{n=1}^N\overline{A}_n(W)\leq M\}. \]If the bandwidth consumed by policy $\pi(W^*)$ satisfies $\sum_{n=1}^N\overline{A}_n(W)=M$, i.e., $\pi(W^*)$ consumes an average bandwidth $M$. Then the optimum solution $\pi_R^*$ to problem 2 is just $\pi(W^*)$. Otherwise, $\pi_R^*$ is a mixture of two policies $\pi_1$ and $\pi_2$, which can be obtained by:
\begin{equation}
	\pi_1=\lim_{W\rightarrow W^-}\pi(W), \pi_2=\lim_{W\rightarrow W^+}\pi(W). \label{eq:policymix}
\end{equation}
To search for policy $\pi(W)$, $\pi_1$ and $\pi_2$, we apply the subgradient descent method. Let $W^{(k)}$ be the Lagrange multiplier used in the $k^{\text{th}}$ iteration. According to \cite[Eq.~6.1.1]{bertsekas2015convex}, the subgradient at $W^{(k)}$ can be computed by:
	\begin{align}
	\text{d}_{W}g( W^{(k)})&=\sum_{n=1}^N\overline{A}_n( W^{(k)})-M.
	\end{align}

We start with $W^{(0)}=0$, if $\sum_{n=1}^N\overline{A}_n(W^{(0)})-M\leq 0$, then scheduling does not have to consider the relaxed bandwidth constraint. The minimum AoI performance to the \textit{RB\&P-Constrained AoI} problem and the lower bound on the AoI performance to the primal \textit{B\&P-Constrained AoI} can be computed simply through:
\begin{equation}
\text{AoI}_{\text{LB}}=\text{AoI}_\text{R}^*=g(0).
\end{equation}

Otherwise, we adopt an iterative algorithm update. By choosing a set of stepsizes $\gamma_k$ similar to \cite{ceran_18_wcnc}, the multiplier for the $k$-th iteration can be computed by:
	\begin{align}
	W^{(k)}=W^{(k-1)}+\gamma_k\text{d}_Wg( W^{(k-1)}).
	\end{align}

The iteration ends until both $|W^{(k)}-W^{(k-1)}|<\varepsilon$  and $\sum_{n=1}^N\overline{A}_n(W^{(k)})\leq M$ are satisfied. Suppose the algorithm terminates at the $K$-th iteration. If $\sum_{n=1}^N\overline{A}_n(W^{(K)})=M$, then $\pi_R^*=\pi(W^{(K)})$. 
Otherwise, we proceed to find two policies $\pi_1$ and $\pi_2$ that constitutes $\pi_R^*$ in Eq.~\eqref{eq:policymix}. Let $W_l$ and $W_u$ be two Lagrange multipliers chosen from sequence $W^{(k)}$, 
\begin{subequations}
	\begin{align}
	W_l\!=\!\arg\max_{W^{(k)}}\sum_{n=1}^N\!\overline{A}_n(\!W^{(k)}\!),\text{s.t. }\sum_{n=1}^N\overline{A}_n(W^{(k)})\!\leq\! M,\\
	W_u\!=\!\arg\min_{W^{(k)}}\sum_{n=1}^N\!\overline{A}_n(\!W^{(k)}\!),\text{s.t. }\sum_{n=1}^N\overline{A}_n(W^{(k)})\!\geq\! M.
	\end{align}
\end{subequations}

Let $M_l=\sum_{n=1}^N\overline{A}_n(W_l)$ and $M_u=\sum_{n=1}^N\overline{A}_n(W_u)$ be the total bandwidth used with respect to minimize the function Eq.~\eqref{relaxedopt}. Suppose $\{\boldsymbol{\mu}^{n, l},\mathbf{y}^{n, l}\}$ is the optimizer to sensor n's LP problem Eq.~\eqref{stationaryLPobj} with multiplier $W_l$ and $\{\boldsymbol{\mu}^{n, u},\mathbf{y}^{n, u}\}$ is the solution with multiplier $W_u$. To satisfy the relaxed bandwidth constraint, the optimum distribution $\{\boldsymbol{\mu}^{n,*}, \mathbf{y}^{n, *}\}$ of the relaxed problem is a linear combination of $\{\boldsymbol{\mu}^{n, l},\mathbf{y}^{n, l}\}$ and $\{\boldsymbol{\mu}^{n, r},\mathbf{y}^{n, r}\}$, which can be computed as follows:
\begin{equation}
	\{\boldsymbol{\mu}^{n,*}, \mathbf{y}^{n, *}\}=\nu\{\boldsymbol{\mu}^{n, l},\mathbf{y}^{n, l}\}+(1-\nu)\{\boldsymbol{\mu}^{n, u},\mathbf{y}^{n, u}\},
\end{equation}
where the mixing coefficient can be computed by:
\[\nu=\frac{M_u-M}{M_u-M_r}.\]
 Consider the structure of each \textit{Decoupled P-Constrained Cost} problem, the optimum scheduling strategy $\pi_R^*$ for the \textit{RB\&P-Constrained} is then constructed as follows:

In each slot $t$, the central controller observe the current AoI $x_n(t)$ and channel state $q_n(t)$ of sensor $n$, a scheduling decision $s_n(t)=1$ is then made with probability $\xi_{x_n(t), q_n(t)}^{n, *}$ is can be computed as follows:
\begin{equation}
	\xi_{x, q}^{n, *}=\begin{cases}
	1, &\xi_{x-1, q}^{n, *}=1\text{ or }\mu_{x,q}^{n, *}=0\text{ or }x\geq X_{\text{max}};\\
	\frac{y_{x, q}^{n, *}}{\mu_{x,q}^{n, *}}, &\text{otherwise}.
	\end{cases}
	\label{SRdistribution}
\end{equation}

The algorithm flow chart to obtain $\xi_{x, q}^{n, *}$ is finally provided as the flow chart Algorithm 1.
\begin{algorithm}
	\caption{Determination of the optimum scheduling probabilities $\xi_{x, q}^{n, *}$ to the \textit{RB\&P-Constrained AoI} Problem}
	\begin{algorithmic}[1]
		\State \textbf{initialization}: start with $W^{(0)}=0$, solve the corresponding LP (20) for each sensor $n$ and compute $\overline{A}_n(W^{(0)})$, denote the optimizer as $\{\boldsymbol{\mu}^{n}, \mathbf{y}^n\}$.
		
		\State{$k\leftarrow 0,M_l\leftarrow 0, M_u\leftarrow2M, \{\boldsymbol{\mu}^{n, l},\mathbf{y}^{n, l}\}=\{\boldsymbol{\mu}^{n, u},\mathbf{y}^{n, u}\}=\{\boldsymbol{\mu}^{n}, \mathbf{y}^n\}$}
		
		\If{$\sum_{n=1}^N\overline{A}_n(W^{(0)})-M\leq 0$}\Comment{Relaxed Bandwidth Constraint is satisfied}
		
		\State{$\{\boldsymbol{\mu}^{n, *},\mathbf{y}^{n, *}\}\leftarrow\{\boldsymbol{\mu}^{n}, \mathbf{y}^n\}, \forall n$}
		
		\Else\Comment{Search for the Lagrange Multiplier}
		\Repeat
		\State{$k\gets k+1$}
		\State{$\text{d}_{W}g( W^{(k-1)})\leftarrow\sum_{n=1}^N\overline{A}_n( W^{(k-1)})-M$}
		\State{$W^{(k)}\gets W^{(k-1)}+\gamma_kd_Wg(W^{(k-1)})$}		\State{Solve the corresponding LP Eq.~\eqref{stationaryLPobj}-\eqref{stationaryLP6} for each sensor $n$ and compute $\overline{A}_n(W^{(k)})$, denote the optimizer as $\{\boldsymbol{\mu}^{n}, \mathbf{y}^n\}$}
		
		\If{$M_l<\sum_{n=1}^N\overline{A}_n(W^{(k)})\leq M$}
		\State{$M_l\gets \sum_{n=1}^N\overline{A}_n(W^{(k)})$}
		\State{$\{\boldsymbol{\mu}^{n, l},\mathbf{y}^{n, l}\}\!\gets\! \{\boldsymbol{\mu}^{n},\mathbf{y}^{n}\}$}
		\ElsIf{$M<\sum_{n=1}^N\overline{A}_n(W^{(k)})\leq M_u$}
		\State{$M_u\gets \sum_{n=1}^N\overline{A}_n(W^{(k)})$}
		\State{$\{\boldsymbol{\mu}^{n, u},\mathbf{y}^{n, u}\}\!\gets\! \{\boldsymbol{\mu}^{n},\mathbf{y}^{n}\}$}
		\EndIf
		\Until{$|W^{(k)}-W^{(k-1)}|<\varepsilon$ and $W^{(k)}\leq M$.}
		\State{$\lambda\gets\frac{M_u-M}{M_u-M_r}$}
		\Comment{Strategy Randomization}
		\State{$\{\boldsymbol{\mu}^{n,*}, \mathbf{y}^{n, *}\}\leftarrow\lambda\{\boldsymbol{\mu}^{n, l},\mathbf{y}^{n, l}\}+(1-\lambda)\{\boldsymbol{\mu}^{n, u},\mathbf{y}^{n, u}\}$}
		\EndIf
		
		\State{Compute $\{\xi_{x, q}^n\}$ according to Eq.~\eqref{SRdistribution}}
		\label{alg:chart}
	\end{algorithmic}
\end{algorithm}

Denote $\text{AoI}_{\text{LB}}$ be the AoI lower bound to the primal \textit{B\&P-Constrained AoI} and let $\text{AoI}_R^*$ be lower bound to the problem \textit{RB\&P-Constrained AoI}. Notice that the AoI performance to the \textit{RB\&P-Constrained AoI} problem can be write out as a function of the  optimizer $\{\boldsymbol{\mu}^{n, *}, \mathbf{y}^{n, *}\}$, and according to the discussion in Section IV-A, the average AoI by following $\pi_R^*$ formulates the lower bound to Problem 1. Hence,
\begin{equation}
\text{AoI}_{\text{LB}}=\text{AoI}_\text{R}^*=\frac{1}{N}\sum_{n=1}^N\sum_{x=1}^{X_\text{max}}\sum_{q=1}^Qx\mu_{x, q}^{n, *}.
\end{equation}
\subsection{Multi-sensor opportunistic scheduling with hard bandwidth constraint}
In this part we construct a truncated policy $\pi$ based on optimal scheduling policy for each of the decoupled sensor and solve the primal \textit{B\&P-Constrained AoI} problem. Let $\pi_R^*$ be the optimum scheduling policy obtained in Section IV(A), where $s_n(t)$ is the scheduling decision under the relaxed constraint, which measures if sensor $n$ is need to be scheduled now. Denote $\Omega(t)=\{n|s_n(t)=1\}$ as the set of sensors that need to be scheduled. The scheduling decision $u_n(t)$ under hard bandwidth constraint is then carried out as follows:
\begin{itemize}
    \item If $|\Omega(t)|\leq M$, i.e., the total number of sensors that currently wait to send updates is less than or equal to the bandwidth resource available, then the scheduling decision $u_n(t)=1, \forall s_n(t)=1$. \item Otherwise if $|\Omega(t)|>M$, the central controller selects a subset of $\mathcal{M}(t)\in\Omega(t), |\mathcal{M}(t)|=M$ sensors from $\Omega(t)$ \textit{randomly} and schedules them to send updates. Those sensors that are in set $\Omega(t)$ but not selected in $\mathcal{M}(t)$ is not scheduled because of limited bandwidth constraint. 
\end{itemize}

\begin{theorem}
	With the proportion of scheduling resources $\frac{M}{N}=\theta$ keeps a constant, the deviation from the optimal scheduling policy for a network with $N$ sensors under the proposed truncated policy $\tilde{\pi}$ is $\mathcal{O}(\frac{1}{\sqrt{N}})$. Thus, with $N\rightarrow\infty$ and $\frac{M}{N}=\theta$, the proposed truncated policy is shown to be asymptotically optimal for the primal \textit{B\&P-Constrained AoI} problem with hard bandwidth constraint.
\end{theorem}
\begin{IEEEproof}
    The detailed proof will be provided in Appendix C.
\end{IEEEproof}

\section{Simulations}
In this section, we provide simulation results to demonstrate the performance of the proposed scheduling policy. We consider a $Q=4$ states channel with the following evolution matrix, where the $j$-th element on the $i$-th row denotes $p_{ij}$, i.e., the probability that channel state evolves from $i$ to $j$:
\[\mathbf{P}=\left[\begin{matrix}
0.4&0.3&0.2&0.1\\0.25&0.3&0.25&0.2\\0.2&0.25&0.3&0.25\\0.1&0.2&0.3&0.4
\end{matrix}\right]. \]
We assume all the sensors have the same above evolving channels and the steady state distribution of channel states is $\boldsymbol\eta=[0.2368, 0.2632, 0.2632, 0.2368]$. The following simulation results are obtained over a consecutive of $T=10^6$ slots.

Notice that from \cite{igor_18_ton}, the optimal policy to minimize AoI performance when all the sensors are identical is a greedy policy that selects the sensor with the largest AoI. If there is no packet-loss in the network, the greedy policy is equivalent to round robin, which requires a minimum power consumption of $\mathcal{E}^{\text{RR}}=\frac{M}{N}\sum_{q=1}^Q\eta_{q}\omega(q)$ for each sensor. In the following simulations, we measure power consumption constraint through ratio $\rho_n=\mathcal{E}_n/\mathcal{E}^{\text{RR}}$. Small $\rho$ indicates that the corresponding sensor has a smaller amount of average power budget. 

\subsection{Average AoI performance}
\begin{figure}[h]
	\centering
	\includegraphics[width=.5\textwidth]{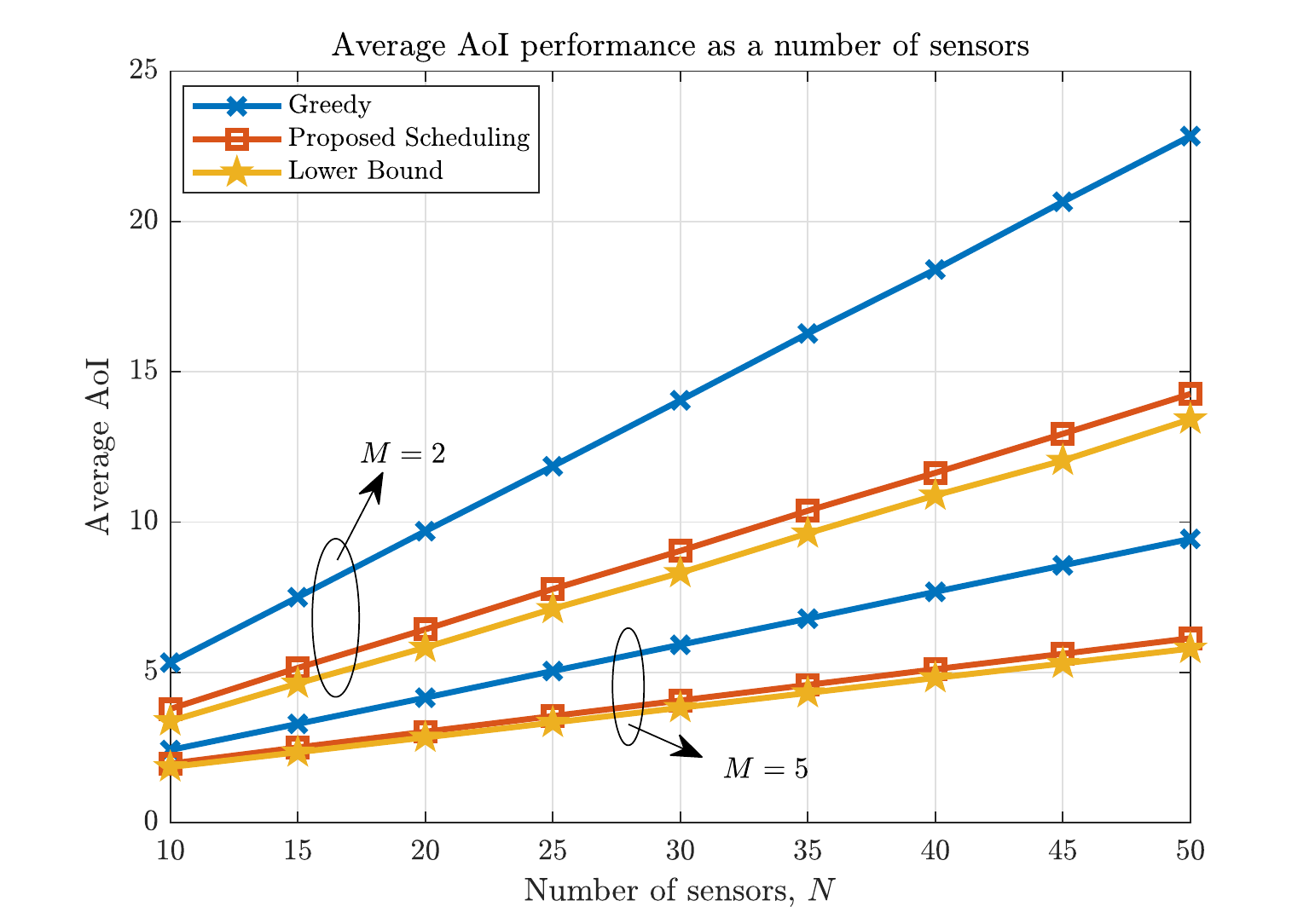}
	\caption{Average AoI performance as a number of sensors $N$, $M=\{2, 5\}$.}
	\label{AoIn}
\end{figure}

Fig. \ref{AoIn} studies average AoI performance as a number of sensors with fixed bandwidth $M=\{2, 5\}$. The power constraint factor is taken from $[0.2, 1.6]$ and $\rho_n=0.2+\frac{1.4}{N-1}(n-1)$. Denote $C_n(t)$ as the total power consumed by sensor $n$ until slot $t$ and let $\mathcal{R}(t)=\{n|\mathcal{E}_nt-C_n(t)\geq0\}$ be the set of sensors that has enough power to support transmission in slot $t$. We compare the proposed policy with a naive greedy policy that selects no more than $M$ sensors with the largest AoI from set $\mathcal{R}(t)$ for scheduling. As can be seen from the figure, the proposed truncated scheduling achieves a close average AoI performance to the lower bound. While the available bandwidth keeps a constant but the number of sensors increases, the proposed truncated policy achieves nearly 40\% average AoI decrease for $M=\{2, 5\}$ in a network with $N=50$ sensors. 

\begin{figure}[h]
	\centering
	\includegraphics[width=.5\textwidth]{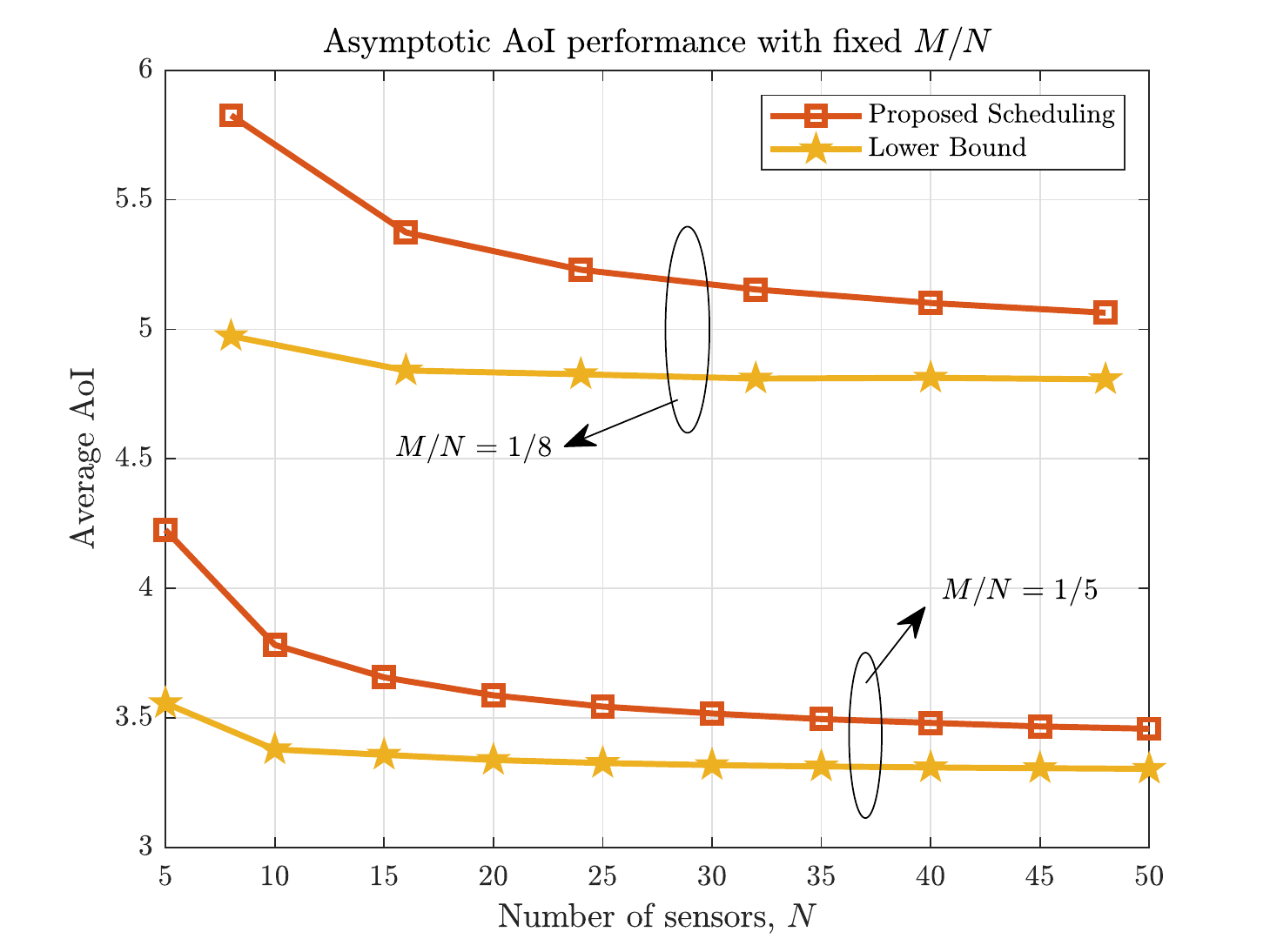}
	\caption{Asymptotic average AoI performance as a number of sensors $N$, available bandwidth is chosen by $M/N=\{\frac{1}{5}, \frac{1}{8}\}$.}
	\label{AoInasymptotic}
\end{figure}

Fig. \ref{AoInasymptotic} studies the asymptotic average AoI performance as a number of sensors, with $\frac{M}{N}=\{\frac{1}{5}, \frac{1}{8}\}$. The power constraint of each sensor is selected by $\rho_n=0.2+\frac{1.4}{N-1}(n-1)$. As can be observed from the figure, the difference between the proposed strategy and the lower bound decreases with $N$. The asymptotic performance is also verified in simulation results.

\subsection{AoI-power trade-off and threshold structure}

\begin{figure}[h]
	\centering
	\includegraphics[width=.5\textwidth]{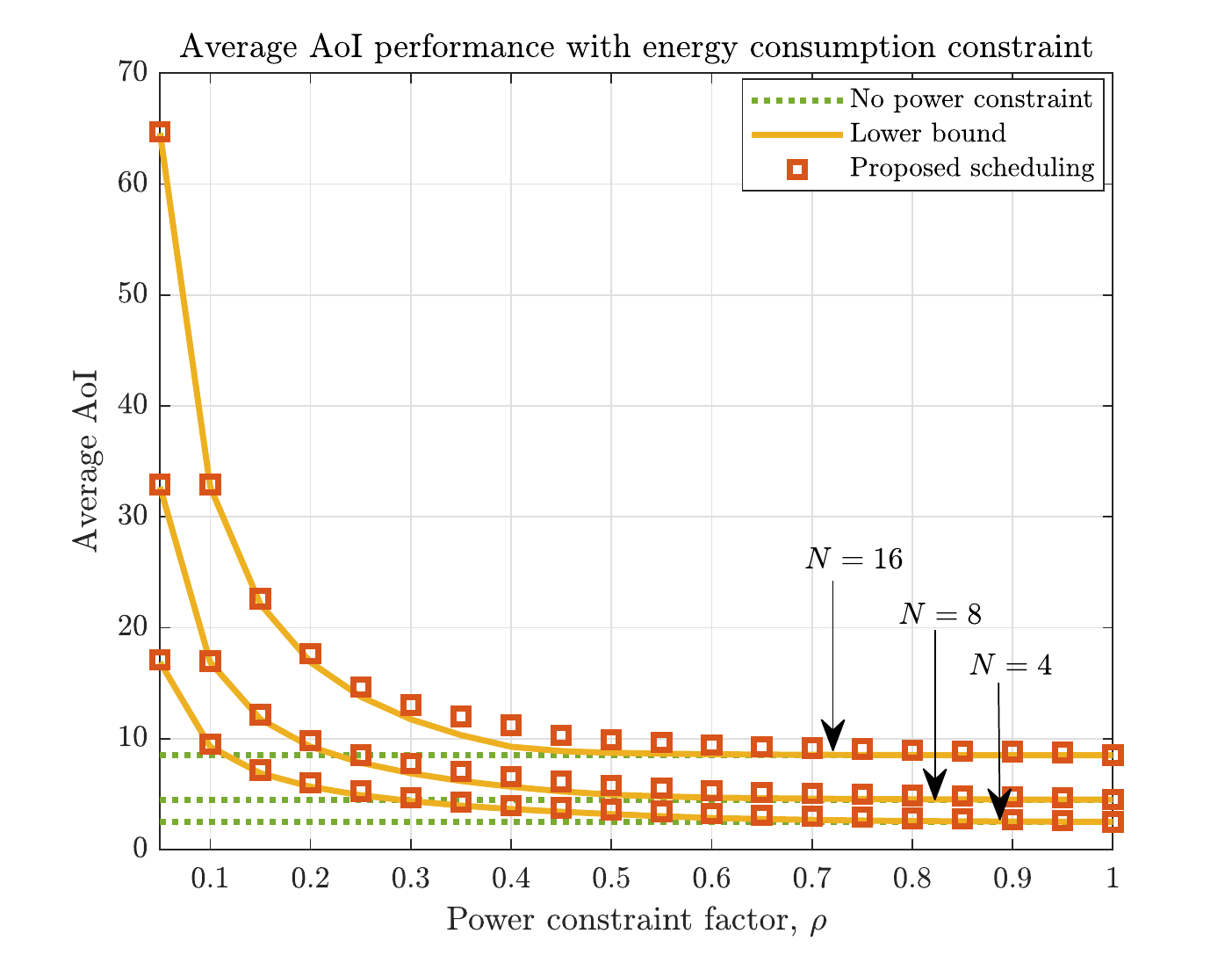}
	\caption{Average AoI-power tradeoff curves for different number of identical sensors.}
	\label{identicalAoI}
\end{figure}

Fig.~\ref{identicalAoI} plots the average AoI-power tradeoff curves for different number of sensors $N=\{4, 8, 16\}$, each sensor has identical channel fading characteristic and the same power constraint factor $\rho$. We assume $M=1$, i.e., only one sensor can be scheduled in each slot. Since all the sensors are identical, it can be concluded that the average scheduling probability of each sensor is smaller than $\frac{1}{N}$. Hence, we can fix $W=0$ and add another constraint on the activation probability to the LP \eqref{LP}, \[\sum_{x=1}^{X_\text{max}}\sum_{q=1}^Qy_{x, q}\leq \frac{1}{N}.\]

By solving this LP problem, we can obtain an lower bound on AoI performance for scheduling multiple identical power constrained sensors. The optimal average AoI performance with no power consumption constraint is plotted in green dashed lines. The yellow solid lines depict AoI obtained by solving the relaxed scheduling problem and red squares represent the AoI performance obtained through the proposed truncated scheduling policy. From the figure, average AoI by following the proposed truncated scheduling policy is close to the AoI lower bound. The average AoI performance decreases monotonically with the power consumption constraint. When $\rho$ is near $1$, indicating each sensor tends to have enough power to carry out a round robin strategy, AoI performance obtained by the proposed truncated scheduling policy and the AoI lower bound also approach the optimal performance by round robin where there is no power constraint. When $\rho$ approaches zero, the average AoI increases dramatically and approaches infinity.

\begin{figure}[h]
	\centering
	\includegraphics[width=.5\textwidth]{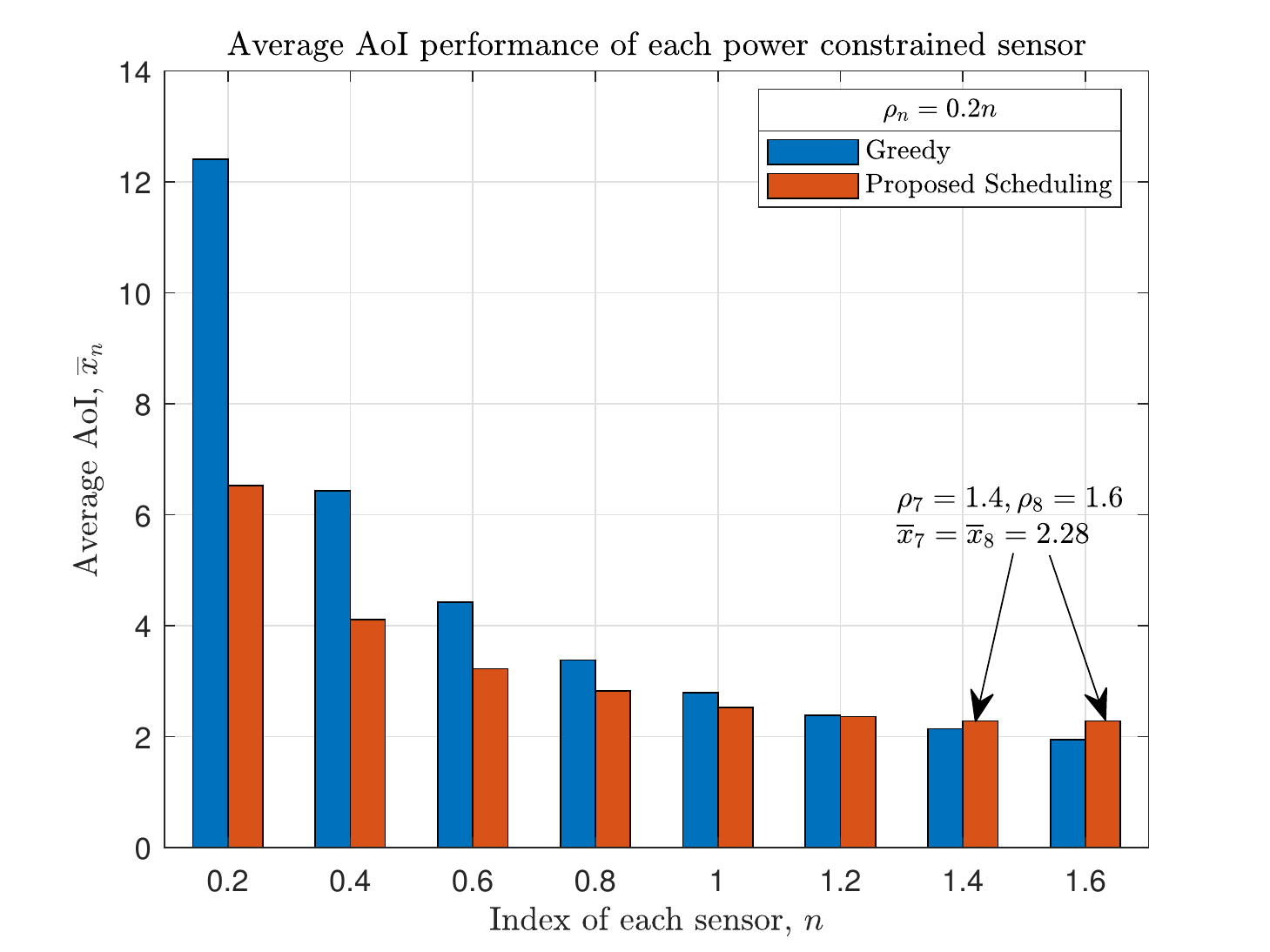}
	\caption{Average AoI performance of each power constrained sensor in a network with $N=8$ sensors and $M=2$, $\rho_n=0.2n$. The average AoI of sensor $n$ is denoted by $\overline{x}_n$.}
	\label{differentenergy}
\end{figure}

\begin{figure*}
	\centering
	\includegraphics[width=\textwidth]{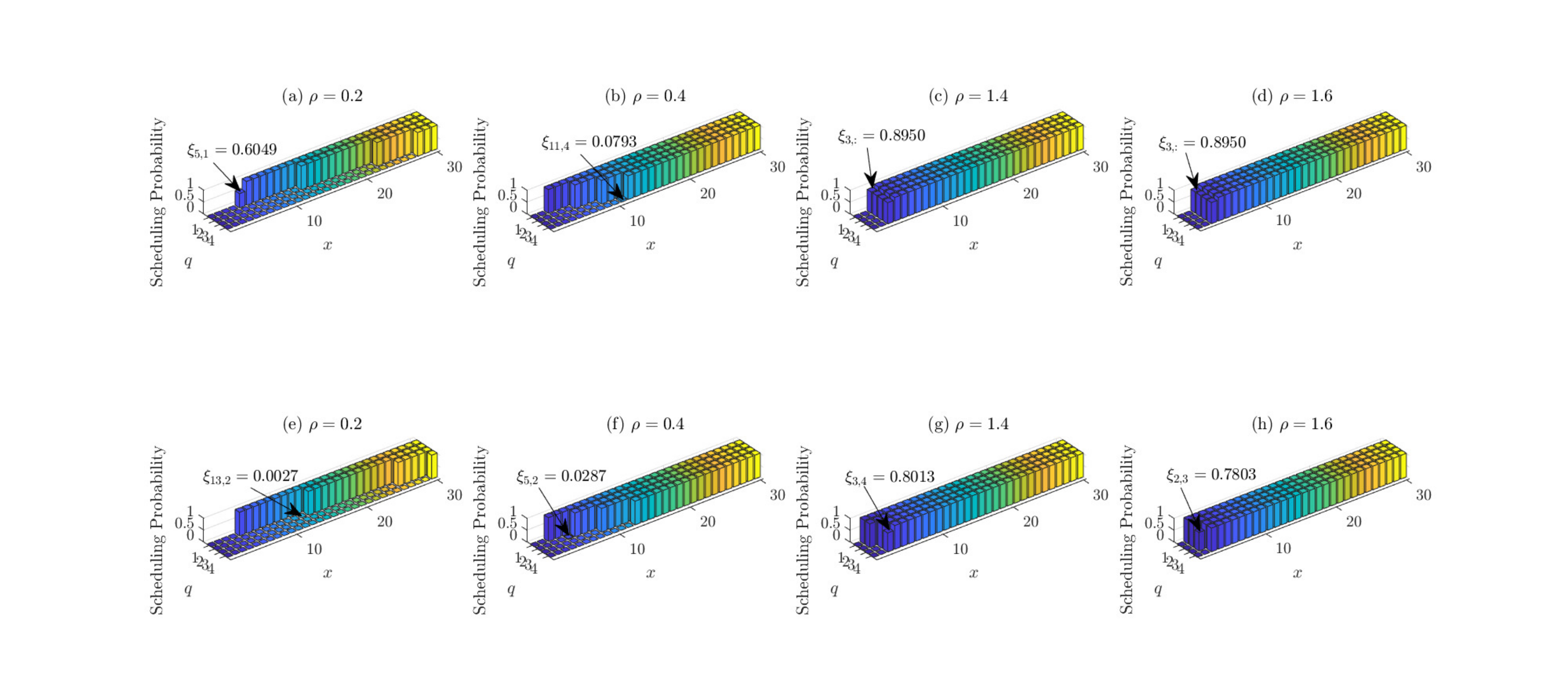}
	\caption{(a)-(d): Scheduling decisions for sensors with different power constraint $\rho_n$ in a network with $N=8$ sensors and $M=2$. (e)-(h): Scheduling decisions for single sensor with different power constraint $\rho_n$ and with no bandwidth constraint. }
	\label{stategyplot}
\end{figure*}

Inspired by the AoI decrease observed in Fig. \ref{AoIn}, we then study the average AoI performance of different power constrained sensor in Fig. \ref{differentenergy} and visualize the scheduling decisions in Fig. \ref{stategyplot}. We consider a network with $N=8$ sensors and $M=2$, each sensor has a power constraint factor $\rho_n=0.2n$. The average AoI performance of sensor $n$ obtained by the proposed algorithm is denoted by $\overline{x}_n$. As is observed from Fig. \ref{differentenergy}, the proposed algorithm brings about 40\% AoI decrease for the first two sensors, which have very limited power for transmission ($\rho_1=0.2, \rho_2=0.4$). The AoI deduction of the proposed algorithm is achieved partly through a more reasonable transmission opportunity allocation to sensors with very limited power. For sensors that have enough power, i.e., sensor 7 and 8 with $\rho_7=1.4, \rho_8=1.6$, our proposed policy guarantees timely updates from those sensors and thus they show similar AoI performance in simulations. 

We visualize the scheduling policy for some representative sensors in Fig. \ref{stategyplot}, where (a)-(d) demonstrate sensor $\{1, 2, 7, 8\}$ with power constraint $\rho=\{0.2, 0.4, 1.4, 1.6\}$, respectively. The optimal scheduling decision for single sensor with power consumption constraint $\rho=\{0.2, 0.4, 1.4, 1.6\}$ but no bandwidth constraint are plotted in (e)-(h). In Fig. \ref{stategyplot}(a) and (b), the transmission power for each sensor is limited, the scheduling threshold $\tau_q$ is an increasing sequence of channel state $q$. Moreover, the threshold of each channel stated in Fig. \ref{stategyplot}(b) is smaller than corresponding threshold in Fig. \ref{stategyplot}(a), indicating that transmission is more likely to happen as a result of more available transmission power. In (a) and (b), the difference between the activation thresholds $\tau_q$ for each sensor is smaller compared with the difference between thresholds illustrated in (e) and (f), indicating the scheduler tries to maintain total probability of sensor scheduling small in order to satisfy the bandwidth constraint of the entire network. Thus, scheduling strategy for a single power constrained sensor seeks to exploit a good channel state, while trying to keep AoI small and use less bandwidth. If unfortunately the channel state is always bad, he will keep waiting until data staleness cannot be bare anymore or the channel state turns good. By comparing Fig. \ref{stategyplot}(a) and (b), the scheduler tries to make full use of the transmission power through a refinement of activation thresholds. \revise{By comparing Fig. \ref{stategyplot}(c) and (g), (d) and (h), when the sensor is equipped with enough power (e.g., $\rho=\{1.4, 1.6\}$), the proposed policy does not use up all the power and all the channel states share the same activation threshold. The threshold is set in order to satisfy the relaxed bandwidth constraint. The bandwidth saved compared with the greedy algorithm is then allocated properly to schedule power constrained sensors and hence achieved significant AoI decrease for those power constrained sensors.} Thus, for a network with different power constrained sensors, the scheduling strategy for different sensors varies according to their power constraints. The scheduler seeks good channels to carry out scheduling decisions for those power constrained sensors, while sensors supported by enough power are updated in a timely manner that can satisfy bandwidth constraint. 

\section{Conclusions}

In this work, we investigate into the problem of age minimization scheduling in power constrained wireless networks, where communication channels are modeled to be an ergodic Markov chain and different level of transmission power is adopted to ensure successful transmission. We decouple the multi-sensor scheduling problem into a single sensor level constrained Markov decision process. We reveal the threshold structure of the optimal stationary randomized policy for the single sensor and convert the optimal scheduling problem into a linear programming. A truncated scheduling policy that satisfies the hard bandwidth constraint is proposed based on the solution to each decoupled sensor. It is revealed that when power of the sensor is very limited, the scheduler seeks to exploit a good channel state while keeping the information fresh. Sensors equipped with enough power are updated in a timely manner that can satisfy the hard bandwidth constraint. 

The network model considered in this work is a very simplified one. In the future, we will extend the work to more general scenarios. Our method generalizes well when the update packet of each sensor arrive stochastically \cite{wangaoi} or packet transmission experiences random packet loss \cite{icc2020}. We will also study scheduling strategy under non-orthogonal multiple access scenario similar to \cite{noma}.
\section*{Acknowledgement}
The authors are grateful to Prof. Philippe Ciblat, Prof. Mich\`ele Wigger from Telecom Paris, Dr. Zhen Zhang,  Mr. Yuchao Chen, Mr. Jingzhou Sun and Mr. Qining Zhang from Tsinghua University, Dr. Bo Zhou from Virginia Tech, Mr. Jiangwei Xu from Wuhan Tech and the anonymous reviewers for helpful suggestions and discussions that greatly improve the presentation and accuracy of the manuscript.
\appendices
\section{Proof of Lemma 1}
\begin{IEEEproof}
	The threshold structure of the optimal policy that minimizes the average cost of \eqref{unconstrainedMDP} is proved by insights from the $\alpha$-discounted cost problems, where $0\!<\!\alpha\!<\!1$ is a discount factor. Given state $(x, q)$, the expected $\alpha$-discounted cost starting from the state over infinite horizons by following policy $\pi$ can be computed:
	\begin{align}
		J_{\alpha,\pi}(x, q)=\lim_{T\rightarrow\infty}\mathbb{E}_\pi\{\sum_{t=0}^T\alpha^t[C_X(x(t), q(t), s(t))\notag\\+\lambda C_Q(x(t), q(t), s(t))]|(x(0)=x, q(0)=q)\}.
	\end{align}
	Let $V_{\alpha}(x, q)=\min_{\pi\in\Pi_{\text{NA}}}J_{\alpha,\pi}(x, q)$ be the minimum expected total discounted cost starting from state $(x, q)$. Then, the minimum total discounted cost will satisfy the following equation:
\begin{align}
V_\alpha(x, q)=\min\{C_X(x, q, 0)+\alpha\sum_{q'\!=\!1}^Qp_{q,q'}V_\alpha(x\!+\!1, q'),\notag\\ C_X(x, q, 1)+\lambda C_Q(x, q, 1)+\alpha \sum_{q'\!=\!1}^Qp_{q,q'}V_\alpha(1, q')\}.
\label{discountBellman}
\end{align}

	To verify the threshold structure of the optimal policy to the total discounted cost problem, we will introduce the following characteristic of $V_{\alpha}(x, q)$:
	
	\begin{lemma}
		For given discount factor $\alpha$ and fixed channel state $q$, the value function $V_\alpha(\cdot, q)$ increases monotonically with $x$. 
	\end{lemma}
	
	The details of the proof will be given in Appendix B. With this lemma, let us now verify the threshold structure. Denote $\Delta(x, q)$ to be the difference in value function by taking $a=\{0, 1\}$, i.e.,
	\begin{align}
		\Delta(x, q)&=C_X(x, q, 0)+\alpha\sum_{q'\!=\!1}^Qp_{q,q'}V_\alpha(x\!+\!1, q')\nonumber\\&-C_X(x, q, 1)-\lambda C_Q(x, q, 1)-\alpha \sum_{q'\!=\!1}^Qp_{q,q'}V_\alpha(1, q')\nonumber\\
		&\!=\!\alpha\sum_{q'\!=\!1}^Qp_{q, q'}(V_\alpha(x\!+\!1, q')\!-\!V_\alpha(1, q'))\!-\!(\lambda\omega(q)\!+\!W).
	\end{align}
	
	Denote $s_\alpha^*(x, q)$ be the optimum solution that achieves the minimum discounted cost $V_\alpha(x, q)$ at state $(x, q)$. If the optimal policy $s_\alpha^*(x, q)=1$, i.e, it is better to schedule the sensor at state $(x, q)$, by substituting Eq.~\eqref{onestepcost} into $\Delta(x, q)\geq 0$, we can obtain the following inequality:
	\begin{equation}
	\alpha\sum_{q'\!=\!1}^Qp_{q,q'}(V_\alpha(x\!+\!1, q')-V_\alpha(1, q'))-(\lambda\omega(q)\!+\!W)\geq 0.
			\label{activatecondition}
	\end{equation}
		According to Lemma 2, the value function $V_\alpha(\cdot, q)$ is monotonic increasing. Hence, for any $x'>x$, $\Delta(x', q)$ can be lower bounded by:
		\begin{align}
		&\Delta(x', q)\nonumber\\
		=&\alpha\sum_{q'=1}^Qp_{q, q'}(V_\alpha(x'+1, q')-V_\alpha(1, q'))-(\lambda\omega(q)+W)\nonumber\\
\overset{(a)}{\geq}&\alpha\sum_{q'=1}^Qp_{q, q'}(V_\alpha(x+1, q')-V_\alpha(1, q'))-(\lambda\omega(q)+W)\geq 0,
		\end{align}
		where inequality (a) is obtained because $V(\cdot, q)$ is increasing. The positivity of $\Delta(x', q)$ implies that for state $x'>x$, the optimal policy for state $(x', q)$ is to schedule the sensor. If at state $(x, q)$ the optimal policy is to be passive, then for state $x'<x$, the optimal policy satisfies $s^*(x', q)=0$ can be verified similarly. 
		
	Moreover, for any state $q$, according to the Bellman equation, the difference between the expected total discounted cost for keeping idle and being scheduled can be computed by
	\begin{align}
		&(C_X(x, q, 0)+\alpha\mathbb{E}[V_\alpha(x+1, q')])\nonumber\\
		&-(C_X(x, q, 1)+\alpha\mathbb{E}[V_\alpha(1, q')])\nonumber\\
		\geq&x+\alpha(x+1)-(x+\alpha\mathbb{E}[V_\alpha(1, q')]+W+\lambda\omega(q))\nonumber\\
		=&\alpha x+\alpha-\alpha\mathbb{E}[V_\alpha(1,q')]-W-\lambda\omega(q),
	\end{align}which increases linearly with $x$. Hence for any channel state, there must be some state such that inequality \eqref{activatecondition} is satisfied. This suggests that the optimal solution cannot keep passive all the time. Thus, there exists a threshold $\tau_q$ for any state $x>\tau_q$, the optimal policy $s_\alpha^*(x, q)=1$ and for state $x<\tau_q$, $s_\alpha^*(x, q)=0$.
	
	Finally, we present the generation of the threshold structure for total discounted cost to establish the structure of the average cost. Take a sequence of discount factors such that $\lim_{k\rightarrow\infty}\alpha_k=1$. Then according to \cite{senottdiscount}, the optimal policy $s_{\alpha_k}^*$ for minimizing the total $\alpha_k$-discounted cost converges to the policy for minimizing the time-average cost, which verifies the threshold structure of the optimal policy $s^*$ as stated in Lemma 1. 
	\end{IEEEproof}

\section{Proof of Lemma 2}
	\begin{IEEEproof}
		In this section, we aim at verifying the monotonic characteristic of the discounted value function. The value of $V_\alpha(x, q)$ can be computed through value iteration regarding the Eq.~\eqref{discountBellman}. Denote $V_\alpha^{(k)}(x, q)$ to be the value function obtained after the $k^{\text{th}}$ iteration, the monotonic characteristic is proved by induction. 

		Suppose $V_\alpha^{(k)}(\cdot, q)$ and $V_\alpha^{(k)}(x, \cdot)$ are non-decreasing. With no loss of generality, suppose $x_1<x_2$. According to the one step cost, we have:
		\begin{equation}C_X(x_1, q, s)<C_X(x_2, q, s), C_Q(x_1, q, s)=C_Q(x_2, q, s).
		\end{equation}
		
		Denote $J_{\alpha,s}^{(k)}(x, q)$ to be the expected total discounted cost if take action $s$ in the $k$-th iteration. Then we have the following inequality:
		 \begin{align}
		 	&J_{\alpha, 0}^{(k)}(x_1, q)\nonumber\\
		 	=&C_X(x_1, q, 0)+\alpha\sum_{q'=1}^Qp_{q,q'}V_\alpha^{(k)}(x_1+1, q')\nonumber\\
		 	\overset{(a)}{<}&C_X(x_2, q, 0)+\alpha\sum_{q'=1}^Qp_{q,q'}V_\alpha^{(k)}(x_2+1, q')\nonumber\\
		 	=&J_{\alpha, 0}^{(k)}(x_2, q),
		 \end{align}
		 where inequality (a) is obtained because of the monotonic characteristic of $V_\alpha^{(k)}(\cdot, q)$. Similarly, we will have the conclusion that $J_{\alpha, 1}^{(k)}(x_1, q)<J_{\alpha, 1}^{(k)}(x_2, q)$. Notice that the value function obtained in the $(k+1)^{\text{th}}$ iteration is obtained by:
		 \[V_\alpha^{(k+1)}(x, q)=\min_{s}J_{\alpha, s}^{(k)}(x, q),\]
		 and for any $s$, $J_{\alpha, s}^{(k)}(x_1, q)<J_{\alpha, s}^{(k)}(x_2, q)$. Thus, the value function $V_{\alpha}^{(k+1)}(x_1, q)<V_{\alpha}^{(k+1)}(x_2, q)$. By letting $k\rightarrow\infty$, the value function $V_\alpha^{(k)}(x, q)\rightarrow V_{\alpha}(x, q)$. Hence, $V_{\alpha}(\cdot, q)$ is monotonic increasing. 
		 
	\end{IEEEproof}

\begin{section}{Proof of Theorem 2}
\begin{IEEEproof}
	Denote $\pi_R^*$ be the policy that in each slot, schedule all the sensors with $s_n(t)=1$ and let $\tilde{\pi}$ be the truncated policy described in Section~V-(B). Since $\pi_R^*$ is the optimum performance to the \textit{RB\&P-Constrained AoI} problem, which formulates the lower bound on the primal \textit{B\&P-Constrained AoI} problem. We verify the asymptotic optimality of the proposed scheduling algorithm by computing the expected AoI difference obtained by $\pi_R^*$ and $\tilde{\pi}$. 
	
	First, considering that $\pi_R^*$ satisfy the relaxed constraint, the average number of sensors that \revise{wait to} send updates by following policy $\pi_R^*$ can then be bounded:
	\begin{equation}
	\overline{\Omega}=\mathbb{E}[|\Omega(t)|]\leq M\label{OmegaUB}.
	\end{equation}
	
According to Lemma 1 and Corollary 2, the optimum policy to each decoupled single-sensor optimization problem possesses a threshold structure. Let $\Gamma_n=\max_q\tau_{n, q}-\min_q\tau_{n,q}$ be the difference between the largest and the smallest scheduling thresholds of sensor $n$ in different channel states. Suppose in slot $t$, $s_n(t)=1$ but sensor $n$ is not scheduled. This phenomenon implies $x_n(t)\geq\min_q\tau_{n, q}$. If the sensor is still not scheduled for $\Gamma_n$ consecutive slots, then its AoI $x_n(t+\Gamma_n)\geq\max_{q}\tau_{n, q}$. Recall that $M/N=\theta$, and the probability that a sensor with $s_n(t)=1$ is not scheduled by policy $\hat{\pi}$ can be computed by $\frac{|\Omega(t)|-M}{|\Omega(t)|}\leq 1-\theta$. Since for $t'\geq t+\Gamma_n$, we have $s_n(t')=1$ and with probability no more than $(1-\theta)$ the sensor $n$ is still not chosen to schedule in slot $t'$. 
Thus the probability that sensor $n$ that should be scheduled in slot $t$ but is not in the next consecutive $t'$ slots is upper bounded by $(1-\theta)^{(t'-\Gamma_n)^+}$, where $(\cdot)^+=\max\{\cdot, 0\}$. Moreover, if the sensor is not scheduled in the consecutive $t'$ slots, policy $\tilde{\pi}$ will cause an extra AoI growth of no more than $t'x_n(t)$ compared with policy $\pi_R^*$. 
	
	Next, we upper bound the effect of truncating in each slot by introducing a modified version of the truncated strategy $\hat{\pi}_R^*$. Based on the relaxed scheduling strategy $\pi_R^*$, when $|\Omega(t)|>M$, the new truncated strategy $\hat{\pi}_R^*$ is designed by: instead of not scheduling a sensor because of limited bandwidth constraint, schedule it as $\pi_R^*$, but add a penalty $\sum_{t'=0}^\infty (1-\theta)^{(t'-\Gamma_n)^+}x_n(t)=(\Gamma_n+\frac{1}{\theta})x_n(t)$ on the total AoI. Notice that the $M$ sensors is chosen randomly, then in slot $t$, if $|\Omega(t)|> M$, the expected extra cost can be upper bounded by:
	\begin{align}
&\mathbbm{1}_{|\Omega(t)|>M}\sum_{n=1}^N\left(\Gamma_n+\frac{1}{\theta}\right)x_n(t)\frac{|\Omega(t)|-M}{|\Omega(t)|}\nonumber\\
\leq&\mathbbm{1}_{|\Omega(t)|>M}\sum_{n=1}^N\left(\Gamma_n+\frac{1}{\theta}\right)x_n(t)\frac{|\Omega(t)|-M}{M}.\end{align}
	otherwise if $|\Omega(t)|\leq M$ there is no extra cost. 
	
	Notice that the AoI obtained by $\hat{\pi}_R^*$ will not decrease compared with $\tilde{\pi}$. Let $x_n(t)$ be the AoI obtained by $\pi_R^*$ and $\mathbbm{1}_{(\cdot)}$ be the indicator function, then the difference between $J(\tilde{\pi})$ and $J(\pi_R^*)$ can be upper bounded as follows:
	\begin{align}
&(J(\tilde{\pi})-J(\pi_R^*))\nonumber\\
\leq&(J(\hat{\pi}_R^*)-J(\pi_R^*))\nonumber\\
=&\frac{1}{NT}\mathbb{E}_{\pi_R^*}\left[\sum_{t=1}^T\mathbbm{1}_{|\Omega(t)|>M}\left(\sum_{n=1}^N\left(\Gamma_n+\frac{1}{\theta}\right)\times\right.\right.\nonumber\\
&\left.\left.\hspace{4.5cm}x_n(t)\frac{|\Omega(t)|-M}{M}\right)\right]\nonumber\\
{=}&\frac{1}{NT}\mathbb{E}_{\pi_R^*}\left[\sum_{t=1}^T\left(\sum_{n=1}^N\left(\Gamma_n+\frac{1}{\theta}\right)\times\right.\right.\nonumber\\
&\hspace{3.5cm}\left.\left.x_n(t)\frac{(|\Omega(t)|-M)^+}{M}\right)\right]\nonumber\\
\leq&\frac{\max_{n}\Gamma_n+\frac{1}{\theta}}{MNT}\mathbb{E}_{\pi_R^*}\left[\sum_{t=1}^T\sum_{n=1}^Nx_n(t)\left(|\Omega(t)|-M\right)^+\right]\nonumber\\
\overset{(a)}{\leq}&\frac{\max_{n}\Gamma_n+\frac{1}{\theta}}{MNT}\mathbb{E}_{\pi_R^*}\left[\sum_{t=1}^T\sum_{n=1}^Nx_n(t)\left(|\Omega(t)|-\overline{\Omega}\right)^+\right]\nonumber\\
\overset{(b)}{\leq}&\frac{\max_{n}\Gamma_n+\frac{1}{\theta}}{MNT}\mathbb{E}_{\pi_R^*}\left[\sum_{t=1}^T\sum_{n=1}^Nx_n(t)\left||\Omega(t)|-\overline{\Omega}\right|\right]\nonumber\\
\overset{(c)}{\leq}&\frac{\max_n\Gamma_n+\frac{1}{\theta}}{MNT}\mathbb{E}_{\pi_R^*}\left[\sum_{t=1}^T\sum_{n=1}^N\max_q\tau_{n, q}||\Omega(t)|-\overline{\Omega}|\right]\nonumber\\
\overset{(d)}{=}&\frac{(\max_n\Gamma_n+\frac{1}{\theta})\sum_{n=1}^N\max_q\tau_{n, q}}{\theta N^2}\times\nonumber\\
&\hspace{3.5cm}\mathbb{E}_{\pi_R^*}\left[\frac{1}{T}\sum_{t=1}^T||\Omega(t)|-\overline{\Omega}|\right],
\label{asymptotic1}
\end{align}
where inequality (a) is because inequality \eqref{OmegaUB} and (b) is because $(\cdot)^+\leq|\cdot|$. Inequality (c) is obtained because following the relaxed strategy $\pi_R^*$, each decoupled sensor has a set of activation thresholds, hence the AoI $x_n(t)$ cannot exceeds the largest thresholds $\max_q\tau_{n, q}$. Equality (d) is because $M=N\theta$.

	Finally, according to \cite{diaconis1991closed}, the expectation of $|\Omega(t)-\overline{\Omega}|$ satisfies:
	\begin{equation}
		\mathbb{E}_{\pi_R^*}[||\Omega(t)|-\overline{\Omega}|]=\mathcal{O}({\sqrt{N}}), \nonumber
	\end{equation}
	which implies:
	\begin{equation}
		\mathbb{E}_{\pi_R^*}\left[\frac{1}{\theta NT}\sum_{t=1}^T||\Omega(t)|-\overline{\Omega}|\right]=\mathcal{O}\left(\frac{1}{\sqrt{N}}\right). \label{asymptotic2}
	\end{equation}
	Notice that the for sensors with fixed power constraint $\mathcal{E}_n$, the difference of threshold structure $\Gamma_n$ does not grow with the number of sensors in the network $N$. In addition, $\frac{M}{N}=\theta$ suggests the available bandwidth $M$ will grow with the number of sensors $N$, thus the thresholds $\max_{q}\tau_{n, q}$ will not grow with $N$. As a result, we will have the following upper bound:
	\begin{equation}
		J(\tilde{\pi})-J(\pi_R^*)=\mathcal{O}\left(\frac{1}{\sqrt{N}}\right).
	\end{equation}
	
	Considering that $J(\pi_R^*)$ is lower bounded by the performance of round robin policy $J(\pi^\text{RR})\geq\frac{1}{2}(\frac{N}{M}+1)$, which has no power consumption constraint. With $\frac{N}{M}=1/\theta$ is a constant and let $N\rightarrow\infty$, we can lower bound $J(\pi_R^*)$ by:
	\begin{equation}
	J(\pi_R^*)\geq J(\pi^{\text{RR}})=\frac{1}{2}\left(\frac{1}{\theta}+1\right).
	\end{equation}
	Finally, the asymptotic optimum performance of the proposed policy $\tilde{\pi}$ can be verified:
	\begin{equation}
		\frac{J(\tilde{\pi})-J(\pi_R^*)}{J(\pi_R^*)}=\mathcal{O}\left(\frac{1}{\sqrt{N}}\right).
	\end{equation}
\end{IEEEproof}
\end{section}
\end{document}